\newcommand\beq{\begin{eqnarray}}
\newcommand\eeq{\end{eqnarray}}
\newcommand\stoponium{\eta_{\tilde t}}

\def\lsim{\mathrel{\rlap{\lower4pt\hbox{$\sim$}}
    \raise1pt\hbox{$<$}}}                % less than or approx. symbol
\def\gsim{\mathrel{\rlap{\lower4pt\hbox{$\sim$}}
    \raise1pt\hbox{$>$}}}

\documentclass[%twocolumn,%showpacs,preprintnumbers,
amsmath,%amssymb,aps,
prd,nofootinbib,floatfix,11pt%,preprint,11pt
]{revtex4}

\allowdisplaybreaks
\usepackage{graphicx}% Include figure files
\usepackage{setspace}

\begin{document}
\renewcommand{\theequation}{\arabic{section}.\arabic{equation}}

\title{\Large%
\baselineskip=21pt
Diphoton decays of stoponium at the Large Hadron Collider}

\author{Stephen P. Martin}
\affiliation{{\it Department of Physics, Northern Illinois University,
DeKalb IL 60115} and\\
{\it Fermi National Accelerator Laboratory,
P.O. Box 500, Batavia IL 60510}}

\begin{abstract}\normalsize \baselineskip=15pt 
If the lighter top squark has no kinematically allowed two-body decays 
that conserve flavor, then it will form hadronic bound states. This is 
required in models that are motivated by the supersymmetric little 
hierarchy problem and obtain the correct thermal relic abundance of dark 
matter by top-squark-mediated neutralino annihilations, or by 
top-squark-neutralino co-annihilations. It is also found in models that 
can accommodate electroweak-scale baryogenesis within minimal 
supersymmetry. I study the prospects for detecting scalar stoponium from 
its diphoton decay mode at the Large Hadron Collider, updating and 
correcting previous work. Under favorable circumstances, this signal will 
be observable over background, enabling a uniquely precise measurement of 
the superpartner masses through a narrow peak in the diphoton invariant 
mass spectrum, limited by statistics and electromagnetic calorimeter 
resolutions.
\end{abstract}

%\pacs{}

\maketitle

%% The arXiv's use of hypertex conflicts with revtex4's use of
%% \tableofcontents in single column format. To avoid this problem,
%% include a file OOREADME.XXX with the word nohypertex in it when
%% you submit to the arXiv. 
\tableofcontents

\vfill\eject
\baselineskip=14.9pt

\setcounter{footnote}{1}
\setcounter{page}{2}
\setcounter{figure}{0}
\setcounter{table}{0}

%%%%%%%%%%%%%%%%%%%%%%%%%%%%%%%%%%%%%%%%%%%%%%%%%%%%%%%%%%%%%%%%%%%%%
\section{Introduction}
\label{sec:intro}
\setcounter{equation}{0}
\setcounter{footnote}{1}

The classic collider signatures for supersymmetry depend on the presence 
of missing energy carried away by a stable, neutral, weakly interacting 
lightest supersymmetric particle in each event. (For reviews, see 
\cite{DGRbook,BTbook,primer}. This paper follows the conventions and 
notations of the last reference.) In a sense, this is disappointing, 
because there are no kinematic mass peaks whose positions would yield 
measurements of superpartner masses. Observables at hadron colliders can 
give precision determinations of superpartner mass differences by way of 
kinematic edges and other distributions, but if $R$-parity is conserved 
the overall mass scale will be much harder to ascertain with precision in 
most models \cite{ATLASTDR,CMSTDR}.

A possible exception occurs if the supersymmetric particles can form 
resonances that annihilate into final states containing only Standard 
Model particles with strong or electromagnetic interactions. An example 
is stoponium, a bound state of a pair of top squarks. The lighter top 
squark (or stop) $\tilde t_1$ has possible flavor-preserving two-body 
decays into the lightest chargino or neutralino,
\beq
\tilde t_1 &\rightarrow& b \tilde C_1,
\label{eq:stoptobot}
\\
\tilde t_1 &\rightarrow& t \tilde N_1,
\label{eq:stoptotop}
\eeq 
which, if kinematically allowed, would cause it to decay long before it 
could form a hadronic bound state. However, the first decay will be 
closed if the chargino $\tilde C_1$ is not at least 5 GeV lighter than 
the top squark, and the large top quark mass means that the second decay 
may well also be kinematically closed. In most of the so-called mSUGRA 
parameter space, this situation is not encountered, but it is 
nevertheless quite possible and even common in other model frameworks. 
Then one must consider three-body (and four-body) decays that preserve 
flavor, and a two-body decay to charm that violates flavor:
\beq
\tilde t_1 &\rightarrow& W^{(*)} b\tilde N_1,
\label{eq:stoptoWb}
\\
\tilde t_1 &\rightarrow& c \tilde N_1 .
\label{eq:stoptocharm}
\eeq
The partial widths associated with the decays (\ref{eq:stoptoWb}) and 
(\ref{eq:stoptocharm}) are known \cite{Hikasa:1987db,Porod:1996at} to be 
far smaller than the binding energy of stoponium or other bound hadronic 
states involving $\tilde t_1$.

Therefore, it is worthwhile on general grounds to consider signals for 
the production and decay of stoponium. At hadron colliders, stoponium is 
produced primarily in gluon-gluon fusion, with the largest cross-section 
for the $1S$ ($J^{PC} = 0^{++}$) scalar ground state, denoted in the 
following as $\stoponium$. This state will decay primarily by 
annihilation, with the possible two-body final states including $gg$, 
$\gamma\gamma$, $W^+W^-$, $ZZ$, $h^0h^0$, $t \bar t$, $b \bar b$, and 
$\tilde N_1 \tilde N_1$. The QCD backgrounds for the gluon and quark 
final states are too huge to contemplate a signal. Also, the $W^+W^-$, 
$ZZ$, and $h^0h^0$ final states are plagued by either large backgrounds 
or small branching fractions, and the $\tilde N_1 \tilde N_1$ final state 
does not give a reconstructable signature. However, Drees and Nojiri in 
\cite{Drees:1993yr,Drees:1993uw} pointed out that $\stoponium \rightarrow 
\gamma\gamma$ provides a viable signal at the CERN Large Hadron Collider. 
(See also refs.~\cite{Nappi:1981ft}-\cite{Inazawa:1993qk} for earlier 
works related to stoponium at hadron colliders.)

Refs. \cite{Drees:1993yr,Drees:1993uw} appeared before two important 
experimental developments which bring this possibility into sharper 
focus. The first is the 1995 discovery of the top quark. Second, the 
results of WMAP and other experiments have bracketed the density of cold 
dark matter in the standard cosmology \cite{WMAP}-\cite{PDG}. This is 
important because the measurement of $\Omega_{\rm DM} h^2 \approx 0.11$ 
can be correlated with a top squark light enough to forbid the decays 
(\ref{eq:stoptobot}) and (\ref{eq:stoptotop}) in at least two scenarios. 
First, neutralino LSP and top-squark co-annihilations can give the 
observed dark matter density if the mass difference $m_{\tilde t_1} - 
m_{\tilde N_1}$ is in a small range 
\cite{stopcoannihilationA}-\cite{stopcoannihilationD}. Second, the 
neutralino LSP can efficiently pair-annihilate into a top-anti-top pair, 
mediated by the $t$-channel exchange of a top squark that is not more 
than about 100 GeV heavier than the LSP 
\cite{Martin:2007gf,Martin:2007hn}. These two scenarios are often 
continuously connected in parameter space, but the latter one requires 
far less fine adjustment of parameters to realize. It does, however, 
require that the gaugino masses are not unified in the way assumed in 
mSUGRA models. A particularly attractive model framework that meets the 
requirements of the stop-mediated annihilation scenario is provided by 
``compressed supersymmetry" \cite{Martin:2007gf,Martin:2007hn}, in which 
the gluino mass parameter is taken to be significantly smaller than the 
wino mass parameter at the scale of apparent gauge coupling unification 
$M_{\rm GUT} \approx 2 \times 10^{16}$ GeV. A reduction of the gluino 
mass compared to the wino and bino mass parameters can also ameliorate 
\cite{KaneKing} the supersymmetric little hierarchy problem. Another 
quite different motivation for a light top squark is provided by models 
that can have a strongly first-order phase transition leading to 
electroweak-scale baryogenesis \cite{baryo}-\cite{baryonew}; this can 
also incorporate the neutralino-stop co-annihilation scenario for dark 
matter.

In this paper, I will consider the $\stoponium \rightarrow \gamma\gamma$ 
signal at the LHC along similar lines to ref.~\cite{Drees:1993uw}, taking 
into account the now known top-quark mass, considering motivated models 
that agree with the observed dark matter density and Higgs mass 
constraints, and using a more liberal angular cut but a more conservative 
energy resolution for the electromagnetic calorimeter. I also correct 
(see Appendix) factor of 2 errors appearing in the $gg$ and 
$\gamma\gamma$ partial decay widths in ref.~\cite{Drees:1993uw}. This 
leads to a somewhat more pessimistic evaluation of the detection 
potential, but in many parts of parameter space the diphoton signal will 
be detectable given a large integrated luminosity at the LHC.

%%%%%%%%%%%%%%%%%%%%%%%%%%%%%%%%%%%%%%%%%%%%%%%%%%%%%%%%%%%%%%%%%%%%%
\section{Diphoton signal for stoponium and backgrounds}
\label{sec:diphotons}
\setcounter{equation}{0}
\setcounter{footnote}{1}

\begin{figure}[!tp]
\begin{minipage}[]{0.4\linewidth}
\caption{\label{fig:binding}
The binding energies for the 1S, 2S, and 1P stoponium states 
as a function of the stoponium mass, as computed from the
potential model of ref.~\cite{Hagiwara:1990sq} with 
$\Lambda^{(4)}_{\rm \overline{MS}} = 300$ MeV.}
\end{minipage}
\begin{minipage}[]{0.57\linewidth}
\includegraphics[width=7.4cm,angle=0]{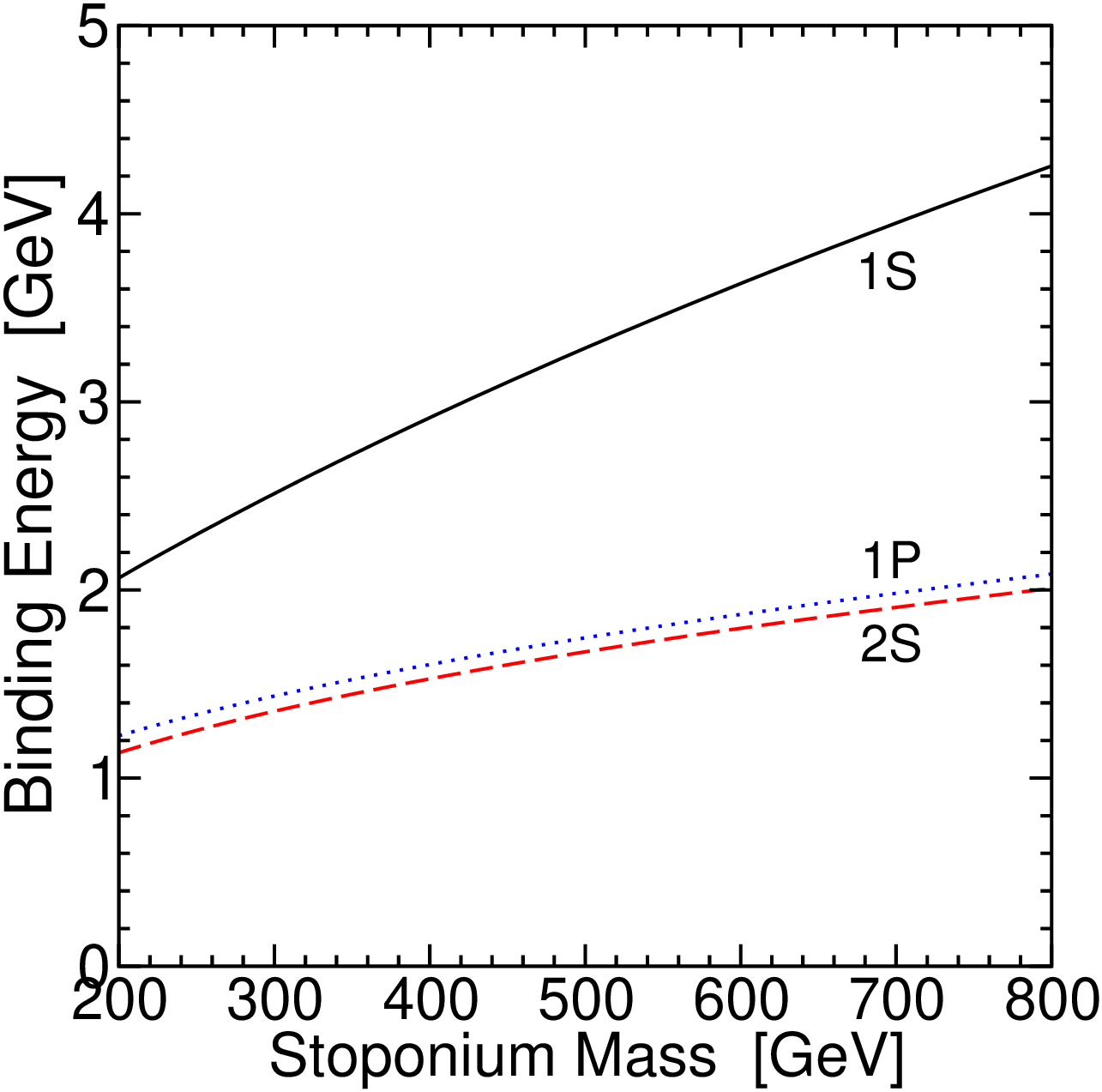}
\end{minipage}
\end{figure}

The leading-order partial decay widths of $\stoponium$ into gluon and 
photon final states are:
\beq
\Gamma(\stoponium \rightarrow gg) &=& \frac{4}{3} \alpha_S^2 
|{R(0)}|^2/{m_{\stoponium}^2},
\\
\Gamma(\stoponium \rightarrow \gamma\gamma) &=& \frac{32}{27} \alpha^2 
|{R(0)}|^2/{m_{\stoponium}^2},
\eeq
where $R(0) = \sqrt{4 \pi} \psi(0)$ is the radial wavefunction at the 
origin. In the Coulomb approximation to the bound state problem, 
$|R(0)|^2/m_{\stoponium}^2 = 4 \alpha_S^3 m_{\stoponium}/27$. However, 
the 
study of ref.~\cite{Hagiwara:1990sq} indicates a softer potential,
with the Coulomb limit not obtained even for very large bound state
masses. In the following, I will adopt the 
$\Lambda^{(4)}_{\rm \overline{MS}} = 300$ MeV
parameterizations of the wavefunction at the origin and the binding
energy as given in 
ref.~\cite{Hagiwara:1990sq}:
\beq
{|R(0)|^2}/{m^2_{\stoponium}} &=& 
(0.1290 + 0.0754 L + 0.0199 L^2  + 0.0010 L^3)\>\,{\rm GeV} ,
\label{eq:Rzero}
\\
 2 m_{\tilde t_1} - m_{\stoponium} 
&=& (3.274 + 1.777 L + 0.560 L^2 + 0.081 L^3)\>\,{\rm GeV} ,
\eeq
where $L = \ln(m_{\tilde t_1}/250\,{\rm GeV})$. The binding energies of 
the 1S ground state and the $2S$ and $1P$ excited states are shown in 
Figure \ref{fig:binding}. It should be noted that these results are based 
on a considerable extrapolation from known experimental results on $c\bar 
c$ and $b \bar b$ bound states, and other potentials can give quite 
different results. (For example, see the ones reviewed in 
ref.~\cite{Barger:1987xg}.) The partial width into gluons is of order 2 
MeV over the considered range of $m_{\stoponium}$. The fact that the 
binding energy is much larger shows that the stoponium bound state will 
indeed form, provided that other partial widths do not overwhelm 
$\Gamma(\stoponium \rightarrow gg)$ by a factor of 1000, a requirement 
easily satisfied by models studied below.

At leading order and in the narrow-width approximation, 
the production cross-section for $\stoponium$ in $pp$ 
collisions is given in terms of its gluonic decay width by
\beq
\sigma (pp \rightarrow \stoponium) = 
\frac{\pi^2}{8 m^3_{\stoponium}} \Gamma (\stoponium \rightarrow gg)
\int_\tau^1 dx \frac{\tau}{x} g(x,Q^2) g(\tau/x, Q^2),
\label{eq:ppstoponium}
\eeq
where $g(x,Q^2)$ is the gluon parton distribution function, and $\tau = 
m_{\stoponium}^2/s$ in terms of the $pp$ collision energy squared $s$. In 
the following, I use the CTEQ5L \cite{Lai:1999wy} set for the parton 
distribution functions, evaluated at $Q = m_{\stoponium}$ for the signal 
and $Q = M_{\gamma\gamma}$ for the backgrounds.

For comparison, the ratio of the stoponium production cross-section to 
that of a Standard Model Higgs boson $H$ with the same mass is just 
$\sigma (pp \rightarrow \stoponium)/\sigma (pp \rightarrow H) = 
\Gamma (\stoponium \rightarrow gg)/\Gamma (H \rightarrow gg)$ at leading 
order. For $m_{\stoponium} = (200, 250, 300, 400, 500, 600, 700, 800)$ 
GeV, this is approximately $\sigma (pp \rightarrow \stoponium)/\sigma (pp 
\rightarrow H) = (1.44, 0.74, 0.40, 0.098, 0.064, 0.054, 0.050, 
0.048)$. However, BR($\stoponium \rightarrow \gamma\gamma$) is much 
larger than BR($H \rightarrow \gamma\gamma$) for masses larger than 140 
GeV, because the latter is loop-suppressed compared to relatively huge 
widths into $ZZ^{(*)}$, $WW^{(*)}$, and $t\overline t$ final states. This 
explains why the stoponium signal $\stoponium \rightarrow \gamma\gamma$ 
can be viable over the mass range where $H\rightarrow \gamma\gamma$ 
observation is not possible. In contrast, ${\rm BR}(\stoponium 
\rightarrow ZZ)/{\rm BR}(H \rightarrow ZZ)$ turns out to be at most about 
$0.3$ over the same mass range. This explains why the search for 
stoponium in $pp \rightarrow \stoponium \rightarrow ZZ$ in the 
$\ell^+\ell^-\ell^{\prime +}\ell^{\prime -}$ and $\ell^+\ell^- 
\nu\overline \nu$ channels is almost certainly not viable 
\cite{Drees:1993uw} for masses where it is the best search option for 
$H$.

Now, multiplying eq.~(\ref{eq:ppstoponium}) 
by the branching fraction into a diphoton state,
and rearranging the factors, one obtains:
\beq
\sigma (pp \rightarrow \stoponium \rightarrow \gamma\gamma) = 
\frac{\pi^2}{8 m^3_{\stoponium}} 
{\rm BR}(\stoponium \rightarrow gg)
\Gamma (\stoponium \rightarrow \gamma\gamma)
\int_\tau^1 dx \frac{\tau}{x} g(x,Q^2) g(\tau/x, Q^2).
\label{eq:sigmappAA}
\eeq
This way of writing the result is useful because, in many realistic 
models, the gluonic decay dominates over all other final states. 
Therefore, it is instructive to adopt an idealized limit where ${\rm 
BR}(\stoponium \rightarrow gg) \approx 1$ as a standard reference 
scenario. (Note that the $\gamma\gamma$ partial width is typically about 
0.005 of the $gg$ partial width.) Then results for particular models can 
be obtained by scaling the signal cross-section by the actual ${\rm 
BR}(\stoponium \rightarrow gg)$.

\begin{figure}[!tp]
\begin{minipage}[]{0.4\linewidth}
\caption{\label{fig:back}
The differential cross-section $d\sigma/dM_{\gamma\gamma}$
for the diphoton backgrounds in $pp$ collisions at
$\sqrt{s} = 14$ TeV, due to the parton-level processes 
$q\bar q \rightarrow \gamma\gamma$ and $gg \rightarrow \gamma\gamma$,
at leading order. Here $M_{\gamma\gamma}$ is the diphoton invariant mass.
The cuts imposed on the angle with respect to the beam axis
are $|\cos\theta_*| < 0.7$
in the diphoton center-of-momentum frame and
$|\cos\theta| < 0.95$ in the lab frame.
The kink in the $gg$ background at $M_{\gamma\gamma} = 2 m_t$ is due to 
the threshold in the top-quark box loop.} 
\end{minipage}
~
\begin{minipage}[]{0.57\linewidth}
\includegraphics[width=8.0cm,angle=0]{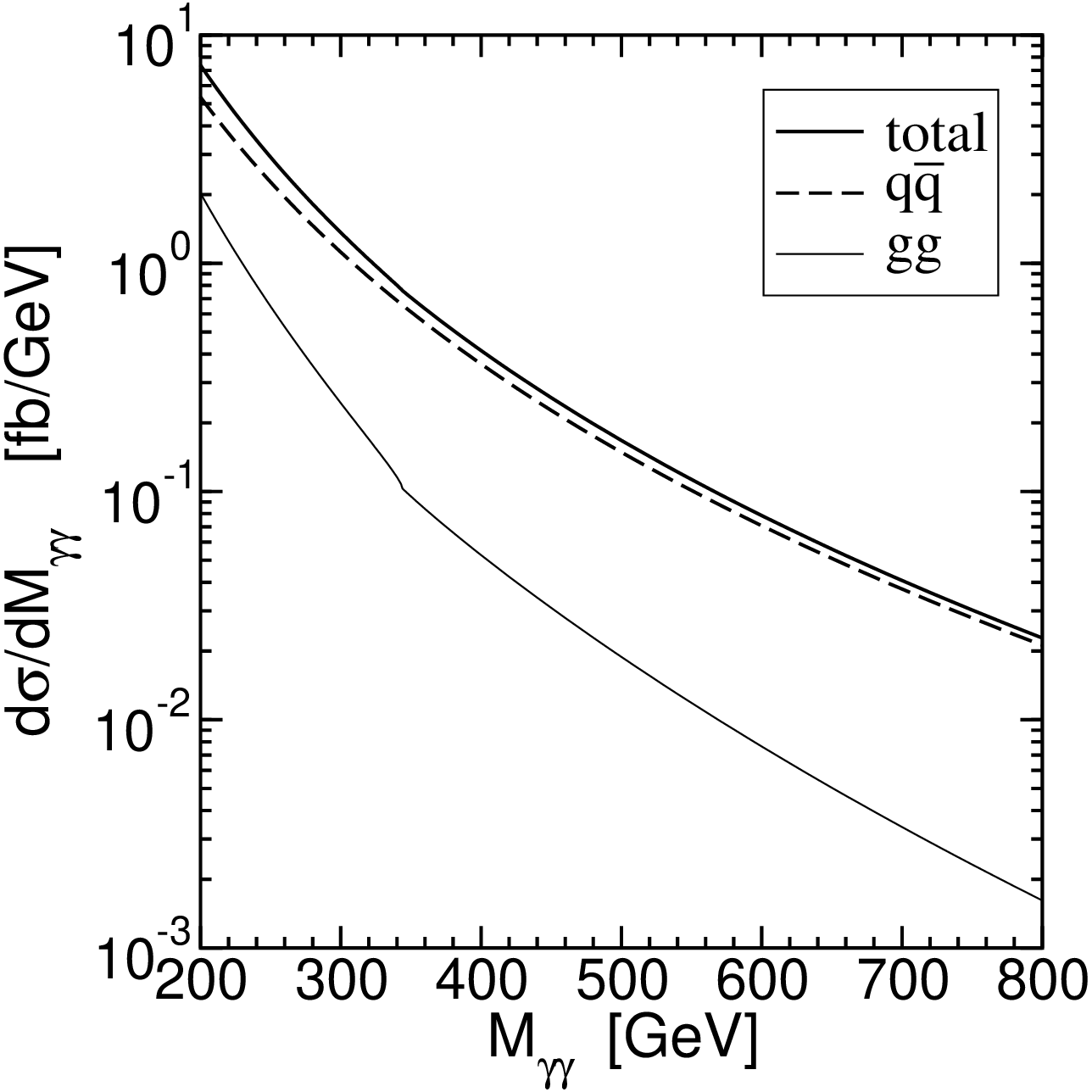}
\end{minipage}
\end{figure}

We next consider the diphoton backgrounds at the LHC. The 
$pp\rightarrow \gamma\gamma$ process has parton-level 
contributions:
\beq
q \overline q &\rightarrow& \gamma\gamma,
\label{eq:qqgammagamma}
\\
g g &\rightarrow & \gamma\gamma,
\label{eq:gggammagamma}
\eeq
with leading-order differential cross-sections found in 
\cite{KarplusNeuman}-\cite{Dicus:1987fk}. The leading order total 
cross-section for eq.~(\ref{eq:qqgammagamma}) is proportional to 
$\ln[(1+z_0)/(1-z_0)] - z_0$, where $z_0$ is the cut on $|\cos\theta_*|$, 
with $\theta_*$ the photon momentum angle with respect to the beam 
direction in the center-of-momentum frame. Since the signal is isotropic, 
with a total signal cross-section proportional to $z_0$, it follows that 
$S/\sqrt{B}$ is maximized for $z_0 \approx 0.705$. The process 
(\ref{eq:gggammagamma}) involves Feynman diagrams that have quark box 
loops. It is somewhat more central than the $q\overline q$ background, 
and so would favor a larger cut $z_0$, but it is quite subdominant over 
the diphoton mass range considered here. For simplicity, I will impose a 
cut on both signal and background of
\beq
|\cos\theta_*| < 0.7
\label{eq:cthetaCMcut}
\eeq
in the center-of-momentum frame. This guarantees a high $p_T$ for the 
photons, for high-mass stoponium states. In addition, I will require that 
the photons be well-separated from the beam direction and the remnant 
beam jets, so
\beq
|\cos\theta| < 0.95
\label{eq:cthetaLABcut}
\eeq
(or $|\eta| < 1.83$)
in the lab frame. The cut (\ref{eq:cthetaCMcut})
is apposite for low rapidities,
and (\ref{eq:cthetaLABcut}) for high rapidities. 
The results for the background at leading order after these cuts are 
shown in fig.~\ref{fig:back} as a function of the invariant mass
of the diphoton system, $M_{\gamma\gamma}$.

Higher-order corrections to the background can be quite important, 
increasing the cross-section after cuts by a factor of 2 or more 
\cite{Balazs:1999yf}-\cite{Balazs:2006cc}. This includes a large 
contribution from the hard scattering process $qg \rightarrow 
\gamma\gamma q$ (and $\overline qg \rightarrow \gamma\gamma \overline q$) 
and the related process $qg \rightarrow \gamma q$ followed by a photon 
from the fragmentation of the quark jet, as well as from double 
fragmentation contributions. Imposing isolation cuts on hadronic activity 
near the photons and requiring the absence of additional hard jets 
reduces these backgrounds considerably. The isolation cut requirement is 
necessary anyway, to eliminate an otherwise large background from jets 
faking photons, including $\pi^0$ and $\eta$ decays that are not resolved 
in the electromagnetic calorimeter. Also, the initial-state gluon 
contributions are not as important in the present case as in the 
well-studied Standard Model Higgs signal mass range $m_H < 140$ GeV, 
because the gluon parton distribution function is relatively smaller at 
larger $x$. The background contributions from jets without charged tracks 
faking photons can likely be reduced to a small level with isolation cuts 
imposed in offline analysis \cite{Lemaire:2006kf}. In any case, because 
the stoponium diphoton resonance will be very narrow, in practice the 
background should be determined directly from LHC data by a sideband 
analysis.

Inclusion of the higher-order background contributions is beyond the 
scope of this paper, especially since the corresponding higher-order 
corrections to the stoponium signal cross-section are not available. In 
addition, the increased background will likely be at least partly 
compensated for by contributions to the signal from production of excited 
stoponium states, followed either by decays to the $1S$ $\stoponium$ 
state by emission of photons or soft mesons or by direct decays to 
$\gamma\gamma$ \cite{Drees:1993uw}. For example, the $2S$ state has a 
binding energy that is probably only slightly less than the $1P$ state 
(see fig.~\ref{fig:binding}), so decays of $2S$ stoponium to the $1P$ 
state and a meson would be kinematically forbidden according to this 
potential model. The $2S$ non-annihilation decays therefore will most 
likely go entirely to the $1S$ ground state. These signal contributions 
will be effectively merged due to the detector energy resolution, leading 
to an overall enhancement of the signal of perhaps a factor of 1.5 (see 
fig.~9 of ref.~\cite{Drees:1993uw}). This unknown enhancement is not 
included here, to be conservative.

\begin{figure}[!tp]
\begin{minipage}[]{0.4\linewidth}
\caption{\label{fig:sigback}
The $pp \rightarrow 
\stoponium \rightarrow \gamma\gamma$ cross-section
and the irreducible
background in a bin with $|M_{\gamma\gamma} - m_{\stoponium}|
< 0.02 m_{\stoponium}$, for $pp$ collisions at
$\sqrt{s} = 14$ TeV, as a function of the stoponium mass
$m_{\stoponium}$. Both are computed at leading order, with 
the same angular cuts
as in Figure \ref{fig:back}. The signal assumes
an idealized limit in which
BR($\stoponium \rightarrow gg$) $+$
BR($\stoponium \rightarrow \gamma\gamma$)
is 100\%.} 
\end{minipage}
\begin{minipage}[]{0.57\linewidth}
\includegraphics[width=8.0cm,angle=0]{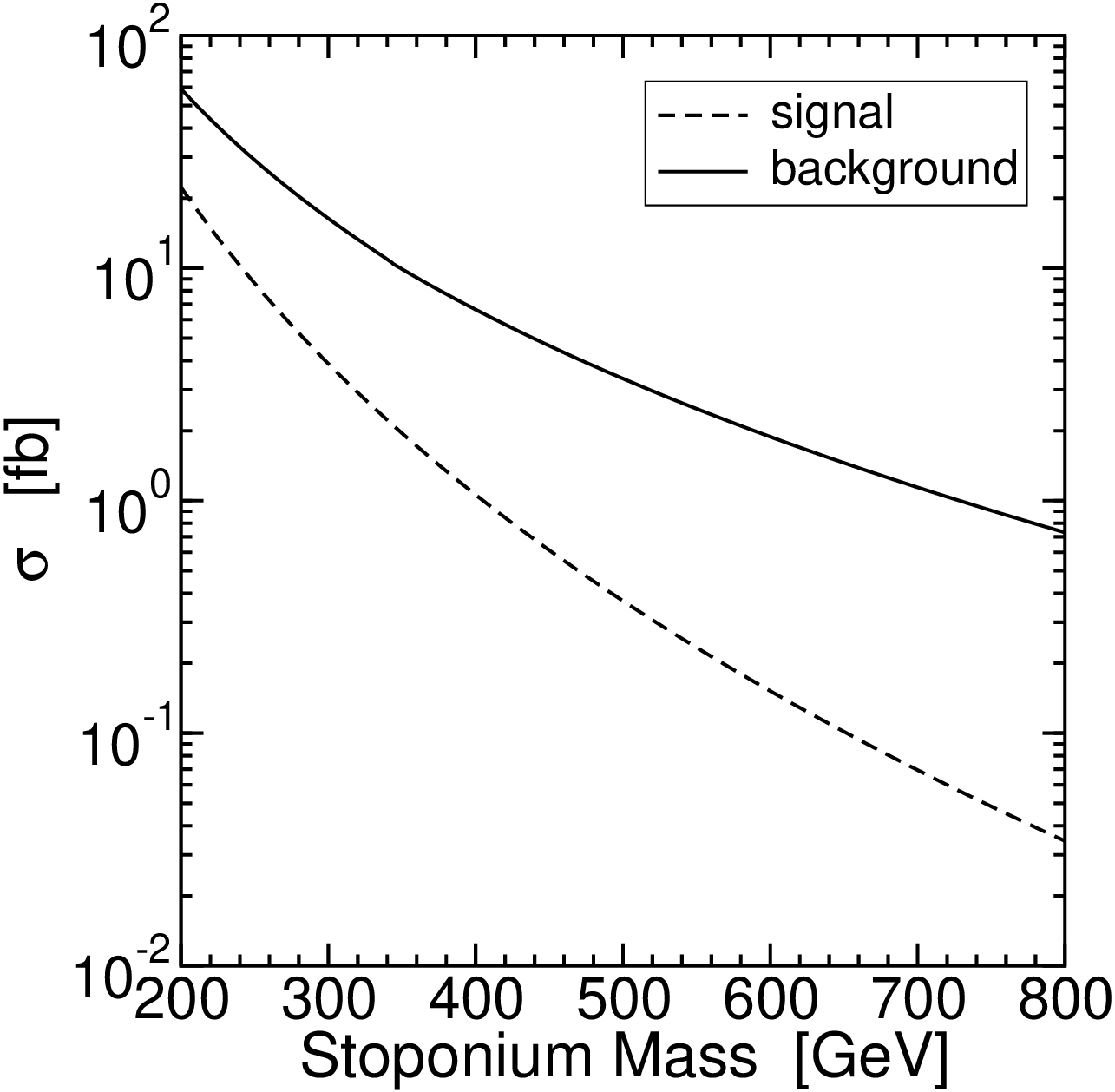}
\end{minipage}
\end{figure}

The energy resolutions of the ATLAS and CMS electromagnetic calorimeters 
at the LHC will clearly dominate over the very small intrinsic width of 
stoponium in determining the experimental width of the signal peak. 
Therefore, to estimate the significance of the signal over the 
background, I consider a bin with width $0.04 m_{\stoponium}$, chosen to 
contain essentially all of the signal peak on which it is centered. (See 
ref.~\cite{Lemaire:2006kf} for a CMS study of a similar diphoton signal 
peak at higher masses, and the CMS and ATLAS physics technical design 
reports \cite{ATLASTDR,CMSTDR} for estimates that put the electromagnetic 
calorimeter resolutions at roughly the per cent level.) The resulting 
comparison of the signal and background within such a bin is found in 
fig.~\ref{fig:sigback}.
Here I have taken the signal corresponding to the idealized reference 
case of ${\rm BR}(\stoponium \rightarrow gg) + {\rm BR}(\stoponium 
\rightarrow \gamma\gamma) = 1$.

\begin{figure}[!tp]
\begin{minipage}[]{0.4\linewidth}
\caption{\label{fig:lumin}
Total integrated luminosity yielding an expected
$S/\sqrt{B} = 2,3,4,5$ for $M_{\gamma\gamma}$ 
in a bin with $|M_{\gamma\gamma} - m_{\stoponium}|
< 0.02 m_{\stoponium}$, as a function of the stoponium mass
$m_{\stoponium}$, for $pp$ collisions at
$\sqrt{s} = 14$ TeV, in the idealized limit that 
BR($\stoponium \rightarrow gg$) $+$
BR($\stoponium \rightarrow \gamma\gamma$)
is 100\%.
The integrated luminosity needed to achieve a given $S/\sqrt{B}$
can be obtained by scaling by 
$1/[{\rm BR}(\stoponium \rightarrow gg)]^2$.}
\end{minipage} 
\begin{minipage}[]{0.57\linewidth}
\includegraphics[width=8.3cm,angle=0]{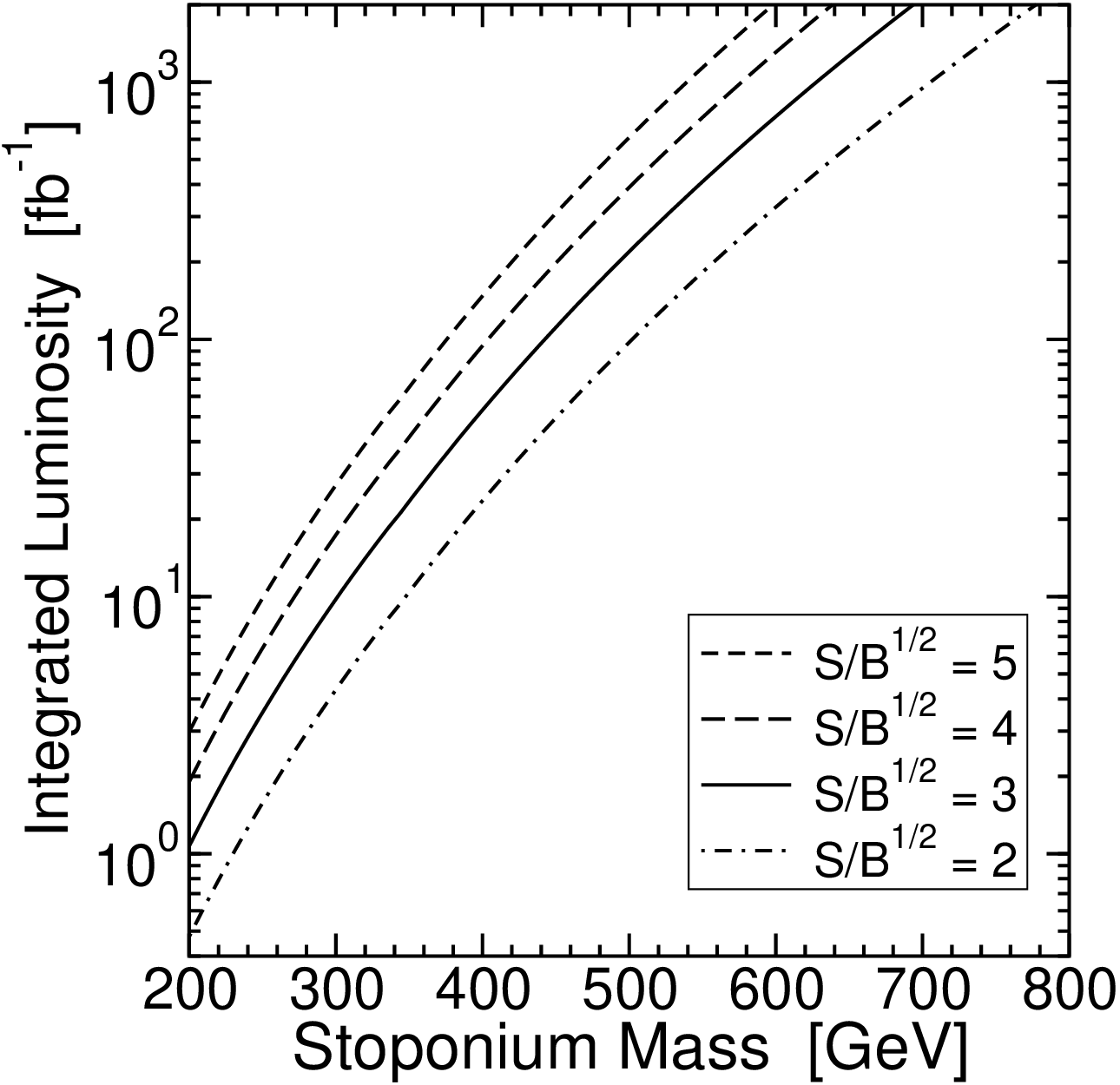}
\end{minipage}
\end{figure}

The resulting integrated luminosities needed to reach significances 
$S/\sqrt{B} = 2,3,4,5$, taking enough events to
use Gaussian statistics, are shown in fig.~\ref{fig:lumin}.
Because there are many bin widths over the mass range considered, the 
probability of a 2-sigma excess in one of them arising just from a 
fluctuation in the background is not negligible. However, the LHC will 
likely have already produced a preliminary estimate of the $\tilde t_1$ 
mass from gluino or direct open stop production before the stoponium 
diphoton signal becomes feasible, so the search range for the stoponium 
mass peak will not be too large. The luminosities in fig.~\ref{fig:lumin} 
should be multiplied by $1/[{\rm BR}(\stoponium \rightarrow gg)]^2$, 
since the required luminosity scales like the square of the signal 
cross-section eq.~(\ref{eq:sigmappAA}).

From fig.~\ref{fig:lumin} one sees that the expected significance for a 
500 GeV stoponium resonance will be at most only about $S/\sqrt{B} = 2$ 
for a canonical high luminosity year of data (100 fb$^{-1}$). This is 
more pessimistic than in ref.~\cite{Drees:1993uw}, due in part to a 
factor of 2 error in that paper in the $\stoponium \rightarrow 
\gamma\gamma$ width, but also due to their assumption of electromagnetic 
calorimeter resolution providing an acceptable bin for $M_{\gamma\gamma}$ 
that is twice as narrow as assumed in the present paper. However, a more 
sophisticated approach based on a maximum likelihood fit, which is beyond 
the scope of this paper, will certainly do better than the simple 
counting in a single bin used here. (Note that the detector mass 
resolution will in any case be smaller than the bin width needed to catch 
all of the signal events.) Also, the integrated luminosity needed is 
proportional to the square of the signal cross-section and to the 
reciprocal of the background, and so is strongly dependent on assumptions 
that are difficult to evaluate confidently at present.
  
%%%%%%%%%%%%%%%%%%%%%%%%%%%%%%%%%%%%%%%%%%%%%%%%%%%%%%%%%%%%%%%%%%%%%
\section{Results for compressed supersymmetry models}
\label{sec:models}
\setcounter{equation}{0}
\setcounter{footnote}{1}

In the previous section, I estimated the integrated luminosity needed to 
achieve detection of the stoponium resonance at a given significance, but 
considering an idealized reference model where BR$(\stoponium \rightarrow 
gg)$ was nearly 100\%. In this section and the next, I consider the 
actual branching ratio achieved in realistic, motivated models that 
satisfy the dark matter density and Higgs mass constraints. As discussed 
in the previous section, the integrated luminosity needed for discovery 
scales like $1/[{\rm BR}(\stoponium \rightarrow gg)]^2$.

First, I will follow ref.~\cite{Martin:2007gf,Martin:2007hn} 
and consider models where the bino, wino,
and gluino masses can be parameterized at $M_{\rm GUT}$ by:
\beq
M_1 &=& m_{1/2} (1 + C_{24}),
\label{eq:binoM}
\\
M_2 &=& m_{1/2} (1 + 3 C_{24}),
\\
M_3 &=& m_{1/2} (1 - 2 C_{24}),
\label{eq:gluinoM}
\eeq
corresponding to an $F$-term source for supersymmetry breaking in a 
linear combination of the singlet and adjoint representations of $SU(5)$. 
For the sake of simplicity, I also assume a common scalar mass $m_0$ and 
scalar trilinear coupling $A_0$ at $M_{\rm GUT}$. For $C_{24}$ of order 
0.2, one finds that the supersymmetric little hierarchy problem is 
ameliorated, with a significant part of parameter space where $\tilde 
t_1$ is the next-to-lightest supersymmetric particle. The dark matter 
thermal relic abundance can be sufficiently suppressed by $\tilde N_1 
\tilde N_1 \rightarrow t \overline t$ due to $t$-channel $\tilde t_1$ 
exchange, giving $\Omega_{\rm DM} h^2 = 0.11$ in accord with 
observations. I impose this as a requirement, by adjusting the value of 
$m_0$ for fixed values of the other parameters, using the program {\tt 
micrOMEGAs 2.0.1} \cite{micrOMEGAs} (checked for approximate agreement 
with DarkSUSY \cite{DarkSUSY}) interfaced to the supersymmetry model 
parameters program {\tt SOFTSUSY 2.0.11} \cite{softsusy} (checked for 
approximate agreement with {\tt SuSpect} \cite{suspect} and ISAJET 
\cite{ISAJET}). The resulting $m_0$ values are reasonably small and do 
not require fine-tuning. In these models, $m_{\tilde t_1} - m_{\tilde 
N_1} < 100$ GeV, so that stoponium will indeed form as a bound state 
before it has a chance to decay.\footnote{The LHC phenomenology (other 
than from stoponium) and dark matter detection prospects of these models 
have been discussed in 
refs.~\cite{Martin:2007gf,Martin:2007hn,Baer:2007uz}. There has recently 
been considerable interest in the phenomenology of other models that 
achieve realistic dark matter with non-universal gaugino masses 
\cite{Corsetti:2000yq}-\cite{Baer:2007xd}.} These stop-mediated 
annihilation models are continuously connected in parameter space to more 
fine-tuned regions in which the $\tilde t_1$, $\tilde N_1$ mass 
difference is just right to allow efficient stop-neutralino 
co-annihilations.

\begin{figure}[!tp]
\includegraphics[width=8.16cm,angle=0]{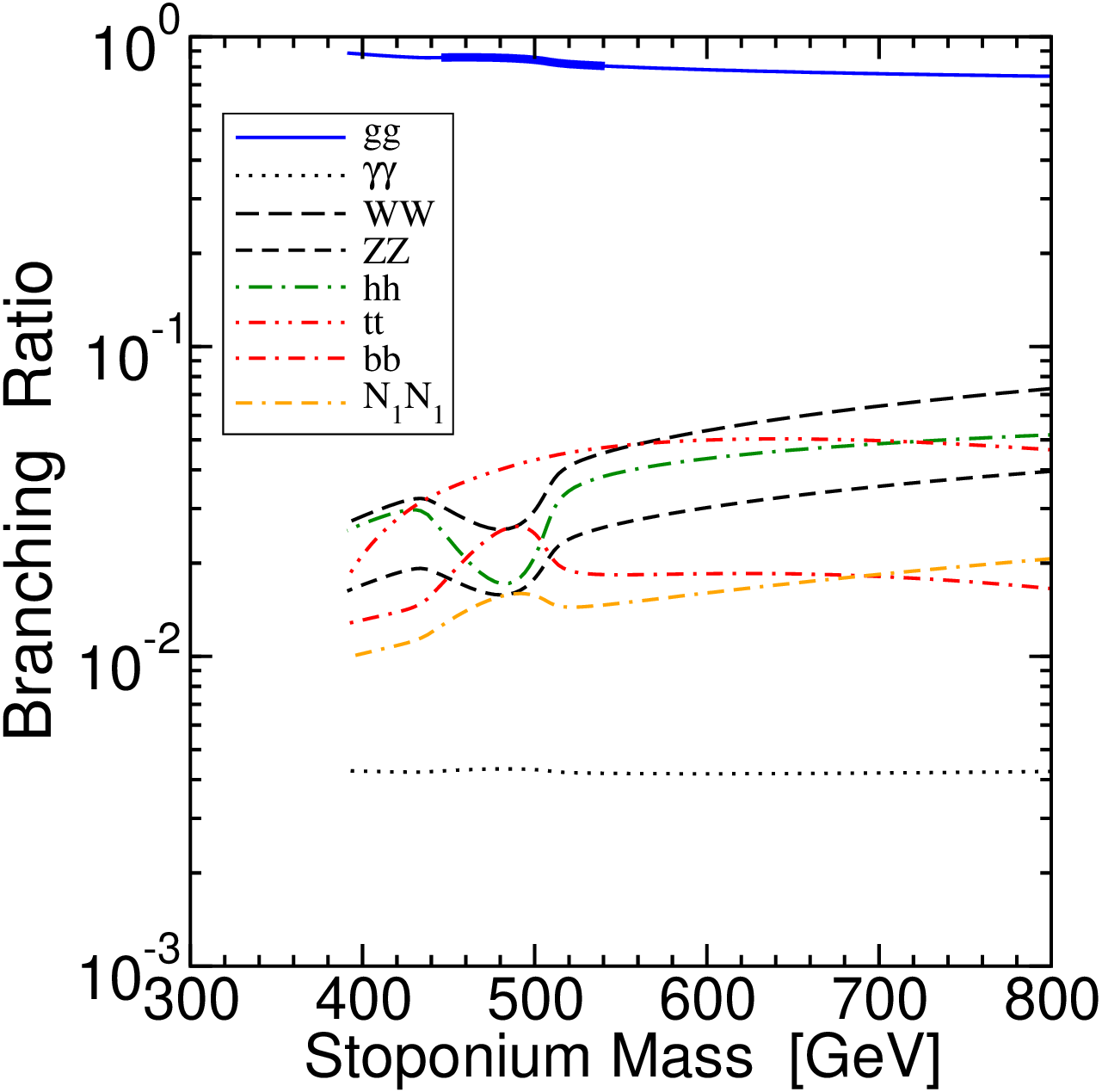}
\includegraphics[width=8.16cm,angle=0]{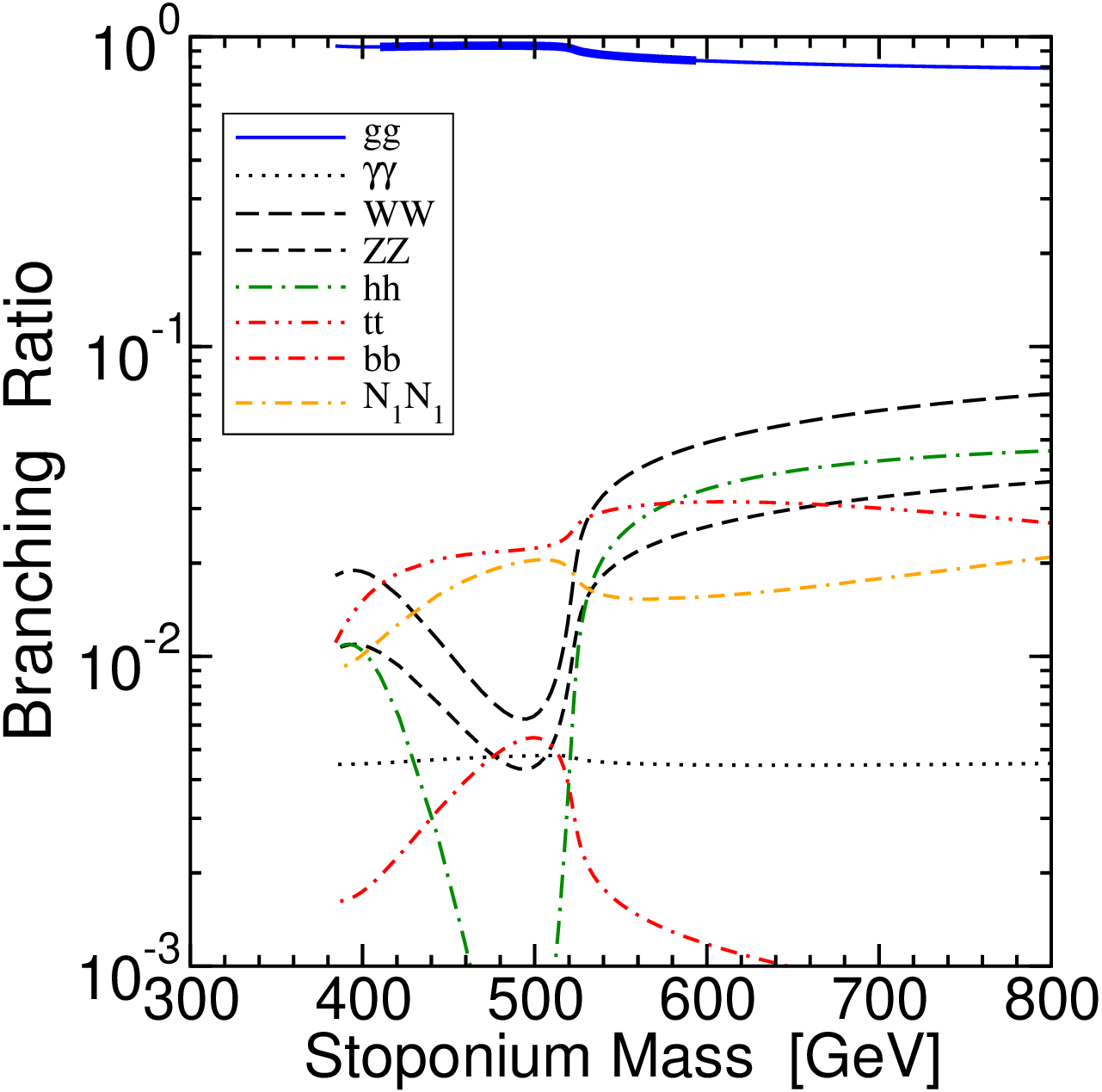}

\includegraphics[width=8.16cm,angle=0]{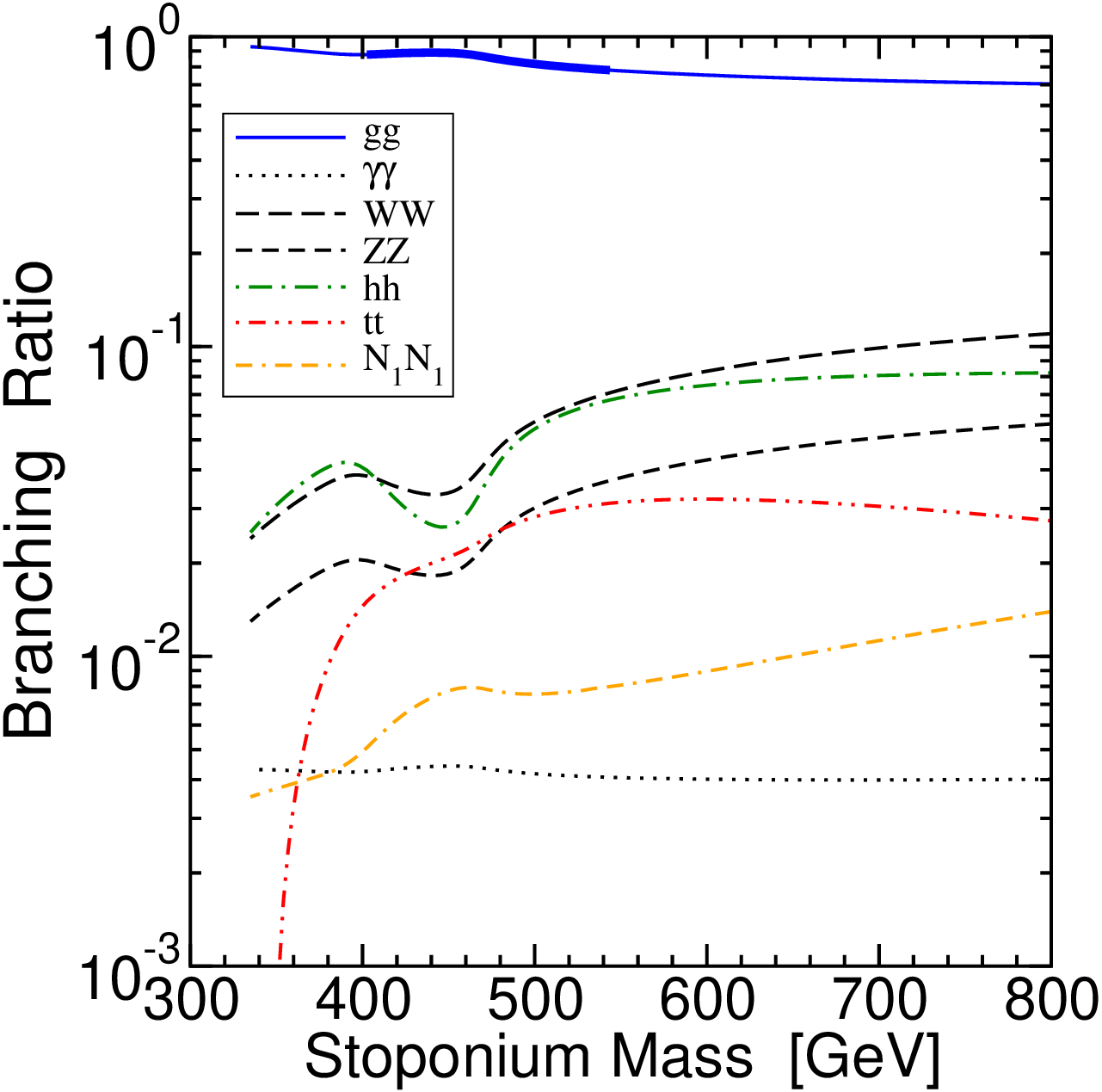}
\includegraphics[width=8.16cm,angle=0]{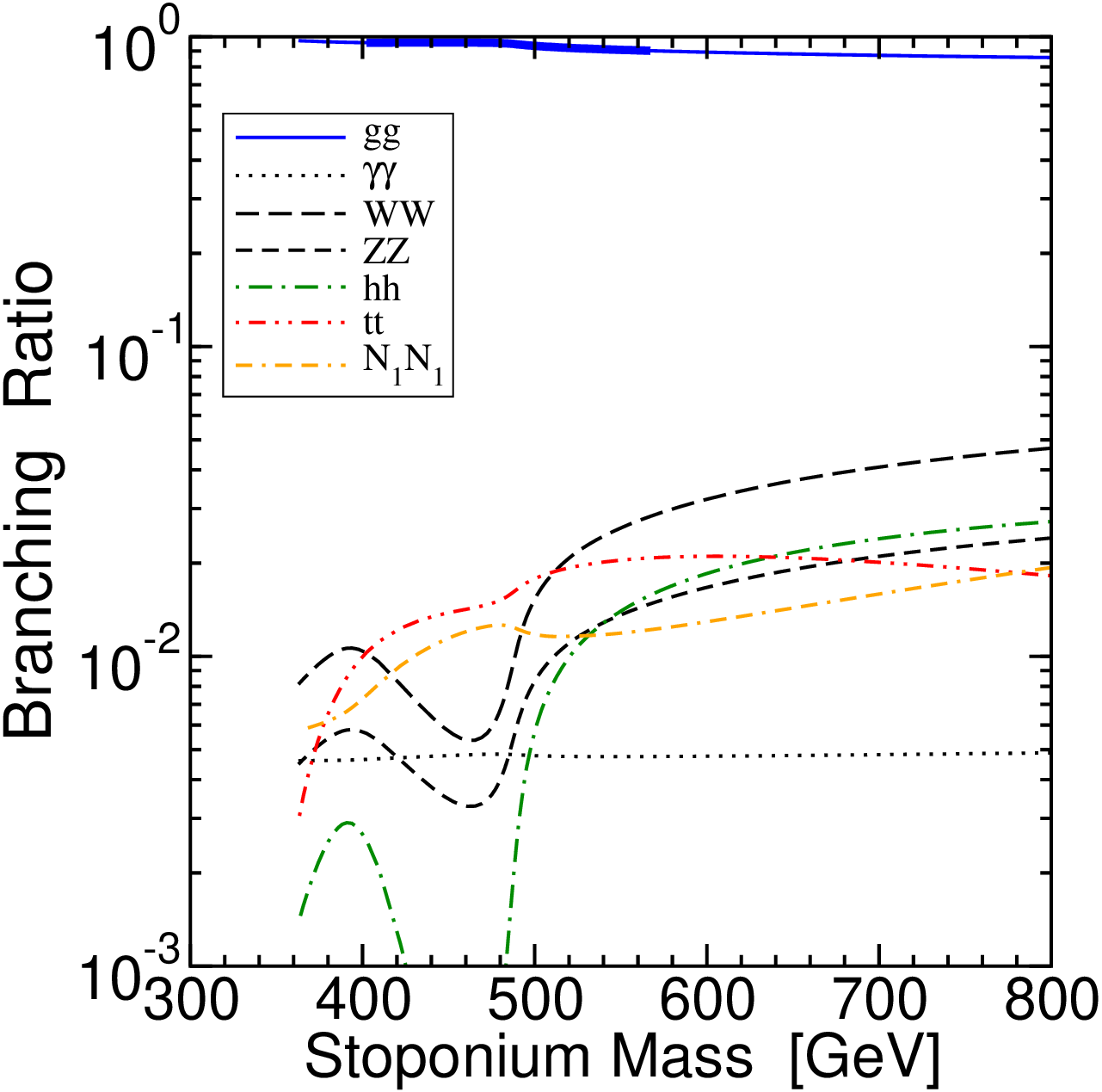}
\caption{\label{fig:BR}
The branching ratios of scalar stoponium into the most important
final states, for some representative models of the type described in
the text. In all cases, $\tan\beta = 10$, $\mu>0$, and $M_1$ varies, with 
$m_0$ adjusted to give $\Omega_{\rm DM} h^2 = 0.11$. The upper left,
upper right, lower left, and lower right panels have respectively
$(C_{24}, -A_0/M_1) = (0.19, 1)$, 
$(0.21, 1)$, $(0.21, 1.5)$, and $(0.24, 1.5)$.
The thicker part of the $gg$ line indicates the range of
stoponium mass for which stop-mediated annihilations 
$\tilde N_1 \tilde N_1 \rightarrow t \overline t$ contribute more 
than 50\% to $1/\Omega_{\rm DM} h^2$.}
\end{figure}

In general, as shown in ref.~\cite{Drees:1993uw}, the most important 
final states
in competition with the $gg$ and $\gamma\gamma$ ones are
$W^+W^-$, $ZZ$, $h^0h^0$, $t \overline t$, $b \overline b$,
and $\tilde N_1 \tilde N_1$.
Formulas for the decay widths for these final states were given in 
ref.~\cite{Drees:1993uw},
and are presented in the Appendix of the present paper in a different
notation. Results for the branching ratios in four typical model lines
are shown in fig. \ref{fig:BR}. These model lines each have a 
continuously
varying overall gaugino mass scale, parameterized by the bino mass parameter
$M_1$, with the wino and gluino masses at $M_{\rm GUT}$ then given by
fixed values of $C_{24}$, as in eqs.~(\ref{eq:binoM})-(\ref{eq:gluinoM}). 
The parameters $\tan\beta = 10$ and $A_0/M_1$ are fixed, and then the
value of $m_0$ is adjusted to give $\Omega_{\rm DM} h^2 = 0.11$.
The four representative model lines were chosen to have
$(C_{24}, -A_0/M_1) = (0.19, 1)$, $(0.21, 1)$, $(0.21, 1.5)$, 
and $(0.24, 1.5)$.
The tuning of $m_0$ needed is particularly mild in the regions indicated 
by the thicker solid (blue) lines for BR$(\stoponium \rightarrow gg)$, 
corresponding to models for which $\tilde N_1 \tilde N_1 \rightarrow t 
\overline t$ dominates the annihilation of dark matter in the early 
universe. This typically gives $m_{\stoponium}$ in the range 400-600 GeV. 
The model lines in fig.~\ref{fig:BR} show the common features that 
BR$(\stoponium \rightarrow gg)$ is quite high, typically 80\% or higher 
for the range shown, and often in excess of 90\% for smaller 
$m_{\stoponium}$ and smaller top-squark mixing. In the mass range where 
neutralino annihilation dominates, the branching ratios for $W^+W^-$, 
$ZZ$, and $h^0h^0$ final states have a welcome dip, due to destructive 
interferences in the amplitudes for each of these final states. The 
dominance of the $gg$ final state yields branching ratio to photons that 
are fairly constant, between 0.004 and 0.005 over the relevant range of 
stoponium masses. Here and in the plots to follow, the lower endpoint on 
the model lines is set by the CERN LEP2 constraint on the Higgs mass, 
taken here to be $m_{h^0} > 113$ GeV due to the theoretical errors in the 
computation.

\begin{figure}[!tp]
\includegraphics[width=8.16cm,angle=0]{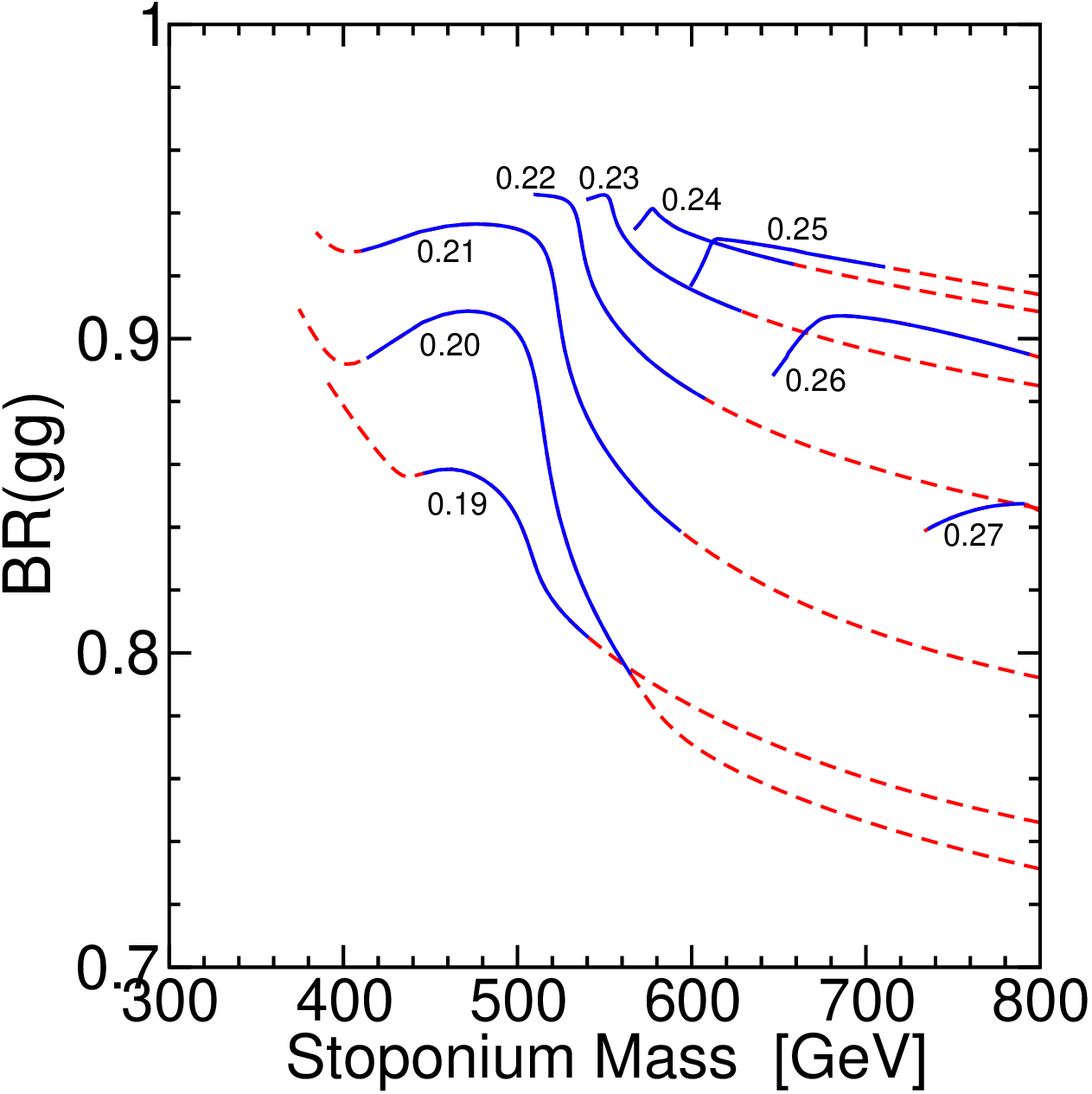}
\includegraphics[width=8.16cm,angle=0]{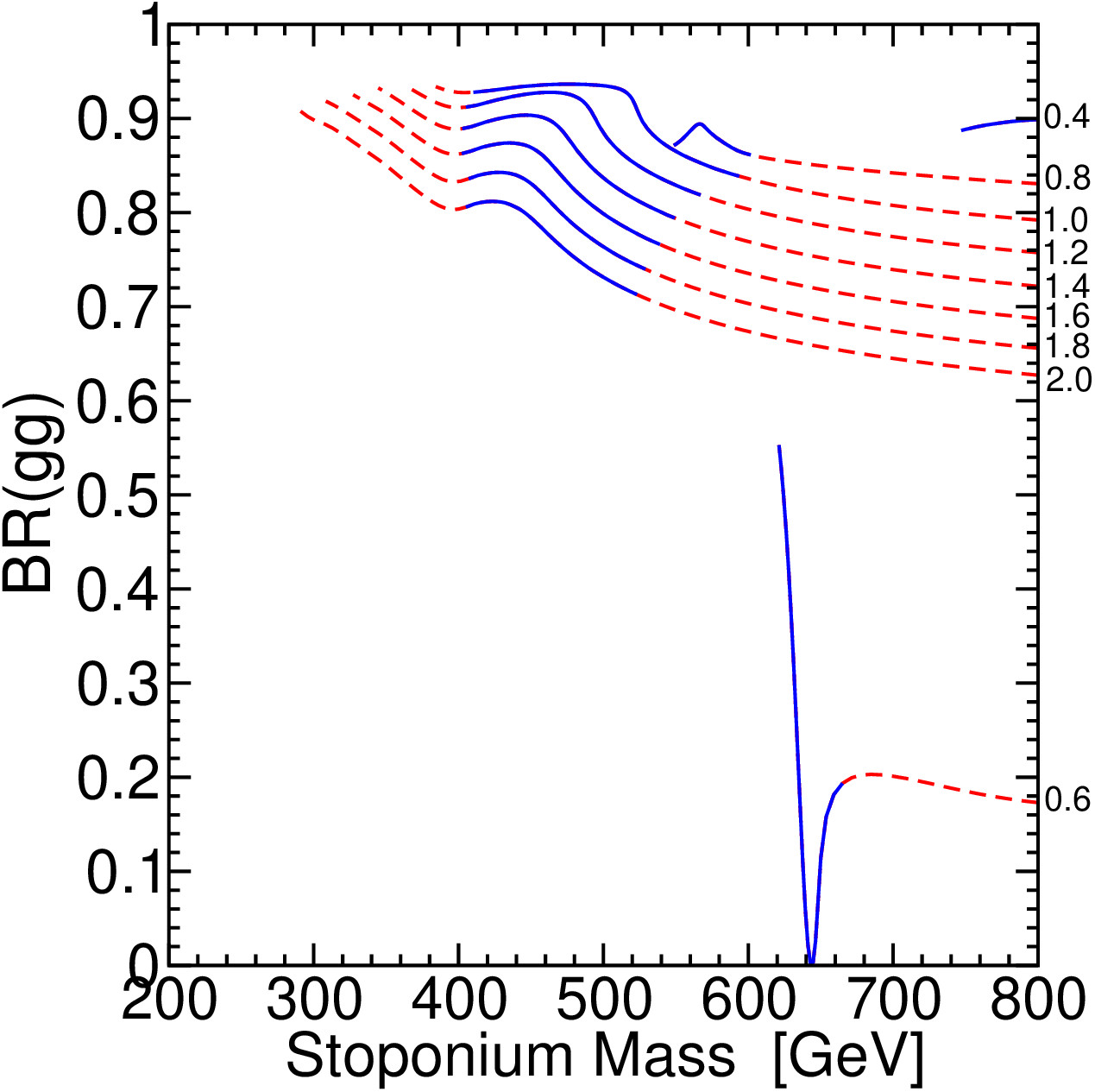}
\caption{\label{fig:BRgg}
The branching ratios of scalar stoponium into the $gg$ final state,
for a variety of models of the type described in the text.
In the left panel, $-A_0/M_1 = 1$ at $M_{\rm GUT}$, with
various $C_{24} = 0.19$ to $0.27$ as labeled.  
In the right panel, $C_{24} = 0.21$, with various $-A_0/M_1 = 0.4$
to $2.0$ as labeled on the far right.
In all cases, $\tan\beta = 10$, $\mu>0$, and $M_1$ varies, with 
$m_0$ adjusted to give $\Omega_{\rm DM} h^2 = 0.11$.
The models for which stop-mediated annihilations 
$\tilde N_1 \tilde N_1 \rightarrow t \overline t$ contribute more 
than 50\% to $1/\Omega_{\rm DM} h^2$ are denoted by solid (blue) lines,
other models by dashed (red) lines.
The integrated luminosity needed to achieve a given $S/\sqrt{B}$
can be obtained from fig.~\ref{fig:lumin} by scaling by 
$1/[{\rm BR}(\stoponium \rightarrow gg)]^2$.}
\end{figure}

The crucial branching ratio into the $gg$ final state is shown in 
fig.~\ref{fig:BRgg} for different slices through parameter space. In the 
left panel, $-A_0/M_1 = 1$ is held fixed and $C_{24}$ is varied over the 
range $0.19$ to $0.27$ for which the stop-mediated neutralino 
annihilation to top quarks mechanism works efficiently, with $m_0$ always 
adjusted to give the observed dark matter density, and $\tan\beta = 10$ 
and $\mu > 0$. Here BR$(\stoponium \rightarrow gg)$ is always high, but 
for larger $C_{24}$, the stoponium mass is forced up by the LEP Higgs 
mass bound and will be difficult or impossible to observe at LHC for 
$C_{24} \gsim 0.26$.

In the right panel of fig.~\ref{fig:BRgg}, we instead fix $C_{24} = 
0.21$, and vary $-A_0/M_1$ in the range from $0.4$ to $2.0$. The 
BR$(\stoponium \rightarrow gg)$ tends to decrease slightly for larger 
$-A_0/M_1$ (as the top-squark mixing increases), leading to enhanced 
branching ratios into $W^+W^-$, $ZZ$, and $h^0h^0$. However, the most 
dramatic effect is seen in the $-A_0/M_1 = 0.6$ case, where the 
BR$(\stoponium \rightarrow gg)$ becomes extremely small, due to the 
effect noted in ref.~\cite{Drees:1993uw} of an $s$-channel resonance in 
stoponium annihilation to $b\overline b$ and $t\overline t$ from $H^0$ 
exchange. This can ruin the possibility of stoponium detection at the 
LHC. In the model framework with a common fixed $m_0$ and other values of 
$C_{24}$, this resonant annihilation to quarks occurs for a range of 
$-A_0/M_1$ less than 1. It is interesting that these models correspond to 
the more optimistic projected sensitivity for the direct detection of 
dark matter in the next generation of low-background underground 
experiments \cite{Martin:2007hn}. However, more generally, the heavy 
neutral Higgs boson mass can be made essentially arbitrary without 
changing the other essential features of the model, by assuming 
non-universal scalar masses at $M_{\rm GUT}$. Therefore, it is impossible 
to make any definitive statements about the complementarity of the 
detectability of the stoponium resonance at LHC and the direction 
detection of dark matter.

\begin{figure}[!tp]
\includegraphics[width=7.9cm,angle=0]{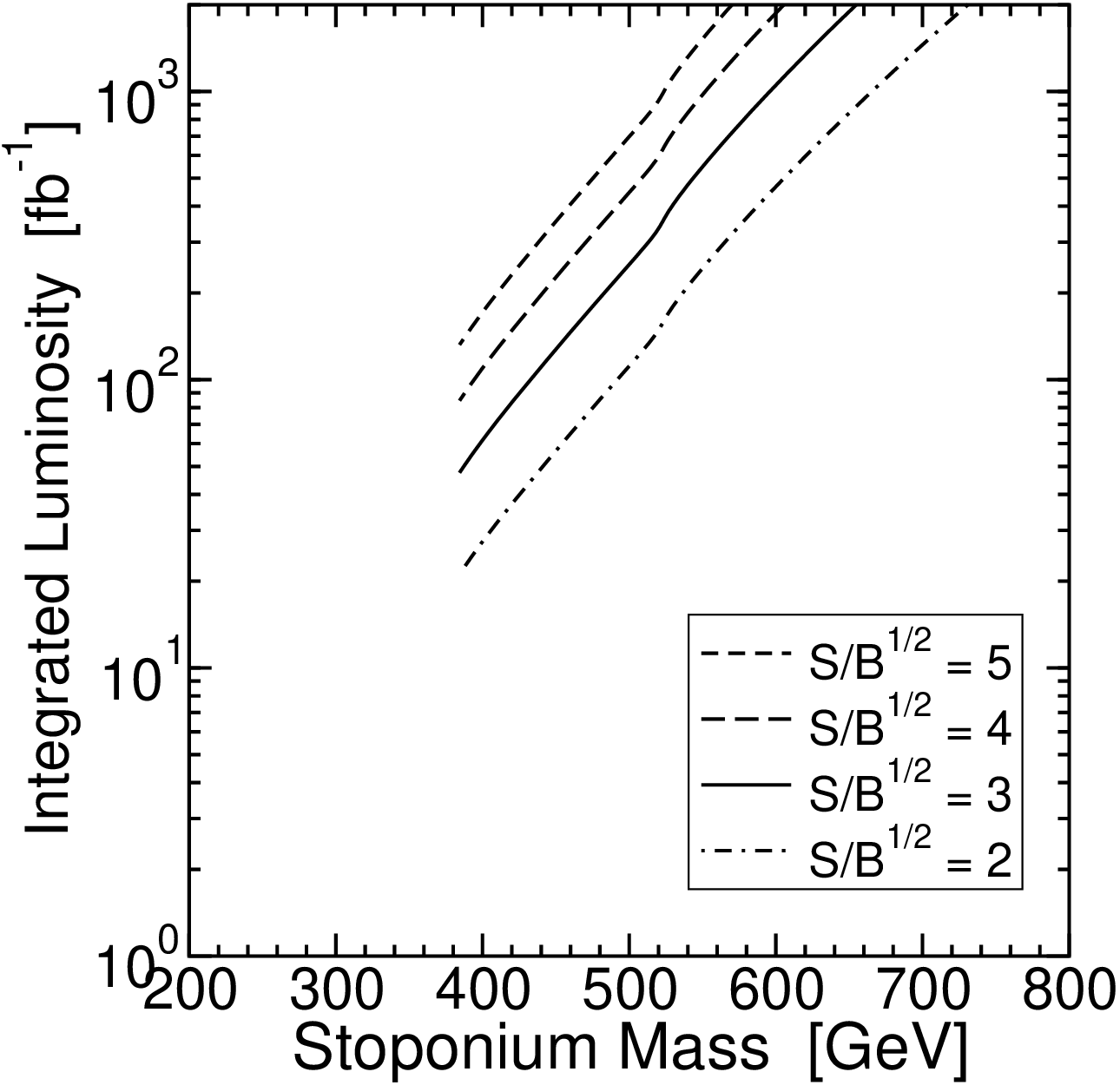}
\includegraphics[width=7.9cm,angle=0]{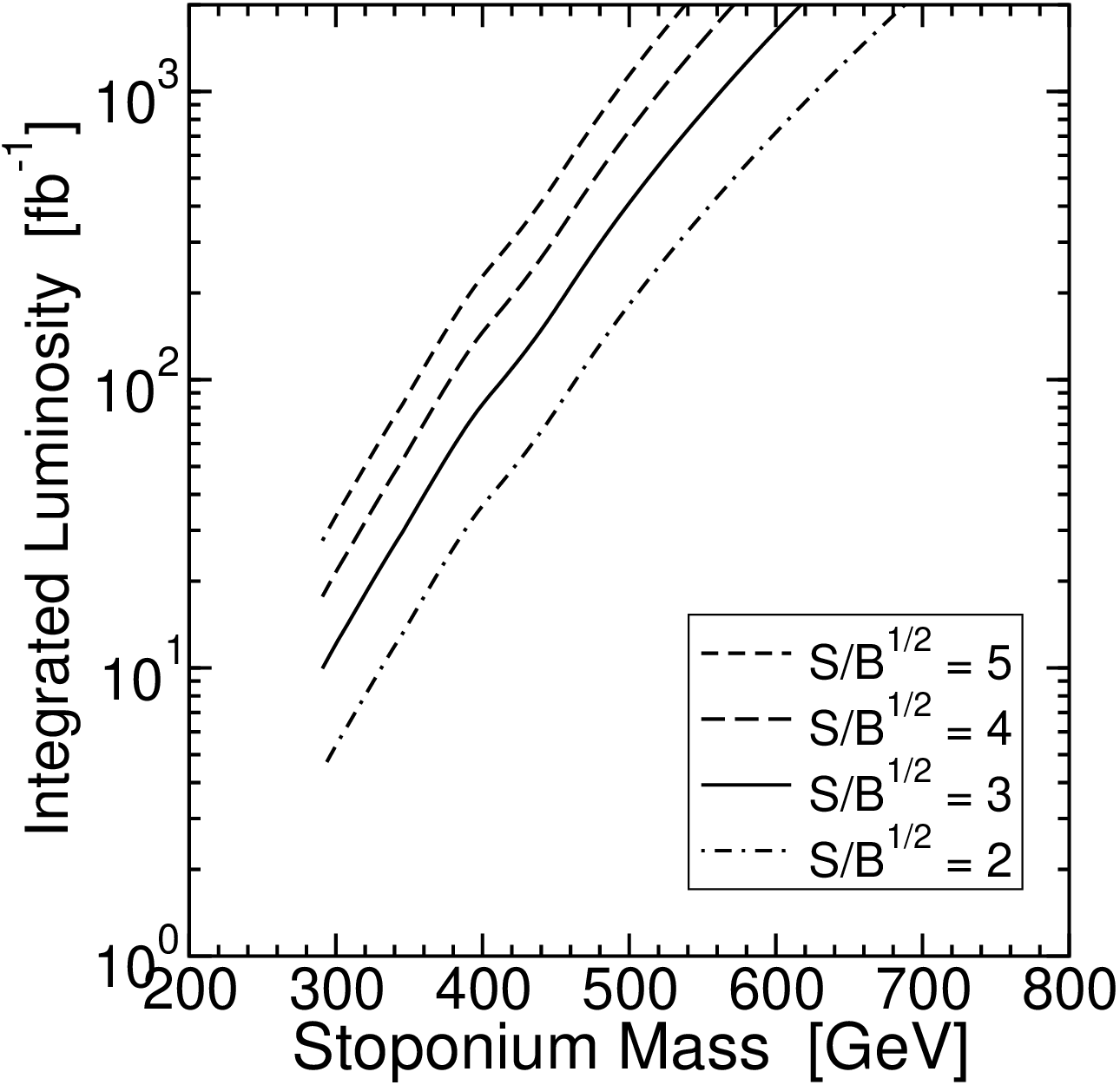}
\caption{\label{fig:lumin_21}
Total integrated luminosity yielding an expected
$S/\sqrt{B} = 2,3,4,5$ for $M_{\gamma\gamma}$ 
in a bin with $|M_{\gamma\gamma} - m_{\stoponium}|
< 0.02 m_{\stoponium}$, as a function of the stoponium mass
$m_{\stoponium}$, for $pp$ collisions at
$\sqrt{s} = 14$ TeV, 
for compressed supersymmetry model lines of the type discussed in the text,
with $C_{24} = 0.21$ and $A_0/M_1 = -1$ (left panel) and
$A_0/M_1 = -2$ (right panel).} 
\end{figure}

Another dangerous decay mode is the $h^0 h^0$ final state, which can 
dominate over all others if $\stoponium \rightarrow h^0h^0$ is not too 
far above threshold, as noted in refs.~\cite{Barger:1988sp,Drees:1993uw}. 
In the dark-matter-motivated models I have studied here, this turns out 
never to be a fatal problem, because the stoponium mass is always 
sufficiently large. This is illustrated in fig.~\ref{fig:lumin_21}, which 
shows the luminosity needed to obtain an expected $S/\sqrt{B} = 2,3,4,5$ 
for $M_{\gamma\gamma}$ in a bin with $|M_{\gamma\gamma} - m_{\stoponium}| 
< 0.02 m_{\stoponium}$, as a function of $m_{\stoponium}$, for two 
representative model lines with $C_{24} = 0.21$ and $A_0/M_1 = -1, -2$. 
If the stoponium mass is small enough, detection might even occur with 
less than $10$ fb$^{-1}$ of data. In general, the models that are 
consistent with the dark matter scenario proposed in 
\cite{Martin:2007gf,Martin:2007hn} can be prime candidates for stoponium 
detection in the diphoton mode, provided that the stoponium mass is not 
too large and not so close to $m_{H^0}$ as to allow a near-resonant 
annihilation decay.

%%%%%%%%%%%%%%%%%%%%%%%%%%%%%%%%%%%%%%%%%%%%%%%%%%%%%%%%%%%%%%%%%%%%%
\section{Results for models with electroweak-scale baryogenesis}
\label{sec:baryo}
\setcounter{equation}{0}
\setcounter{footnote}{1}

Another motivation for a relatively light top squark is the possibility 
of achieving electroweak-scale baryogenesis in the MSSM 
\cite{baryo}-\cite{baryonew}. The necessity of a strongly first-order 
phase transition to a meta-stable electroweak symmetry-breaking vacuum 
limits the allowed parameter space, requiring a mostly right-handed top 
squark with mass less than $m_t$. (For more details, see 
\cite{baryo}-\cite{baryonew}.) Here, I will consider a model framework 
proposed in \cite{baryoDM}, with an off-diagonal top-squark squared mass 
matrix element $m_t X_t$ with $0.3 \lsim |X_t|/m_{\tilde t_2} \lsim 0.5 
$, and $m_{\tilde t_2}$ very large (here 10 TeV), $\tan\beta = 5$ to 10, 
$m_{h^0}$ between 115 and 120 GeV, and all other superpartners except the 
LSP supposed to be sufficiently heavy that they do not mediate large 
contributions to the stoponium decay width. Then the parameters with the 
most important impact on the stoponium decay widths are $m_{h^0}$, 
$m_{\tilde t_1}$, and the top-squark mixing angle. The region of 
parameter space where electroweak-scale baryogenesis can work is roughly 
120 GeV $< m_{\tilde t_1} < $ 135 GeV \cite{baryonew}, but I will 
consider a wider range consistent with a meta-stable vacuum and the Higgs 
mass constraint from LEP2 \cite{baryoDM}.

\begin{figure}[!tp]
\includegraphics[width=8.0cm,angle=0]{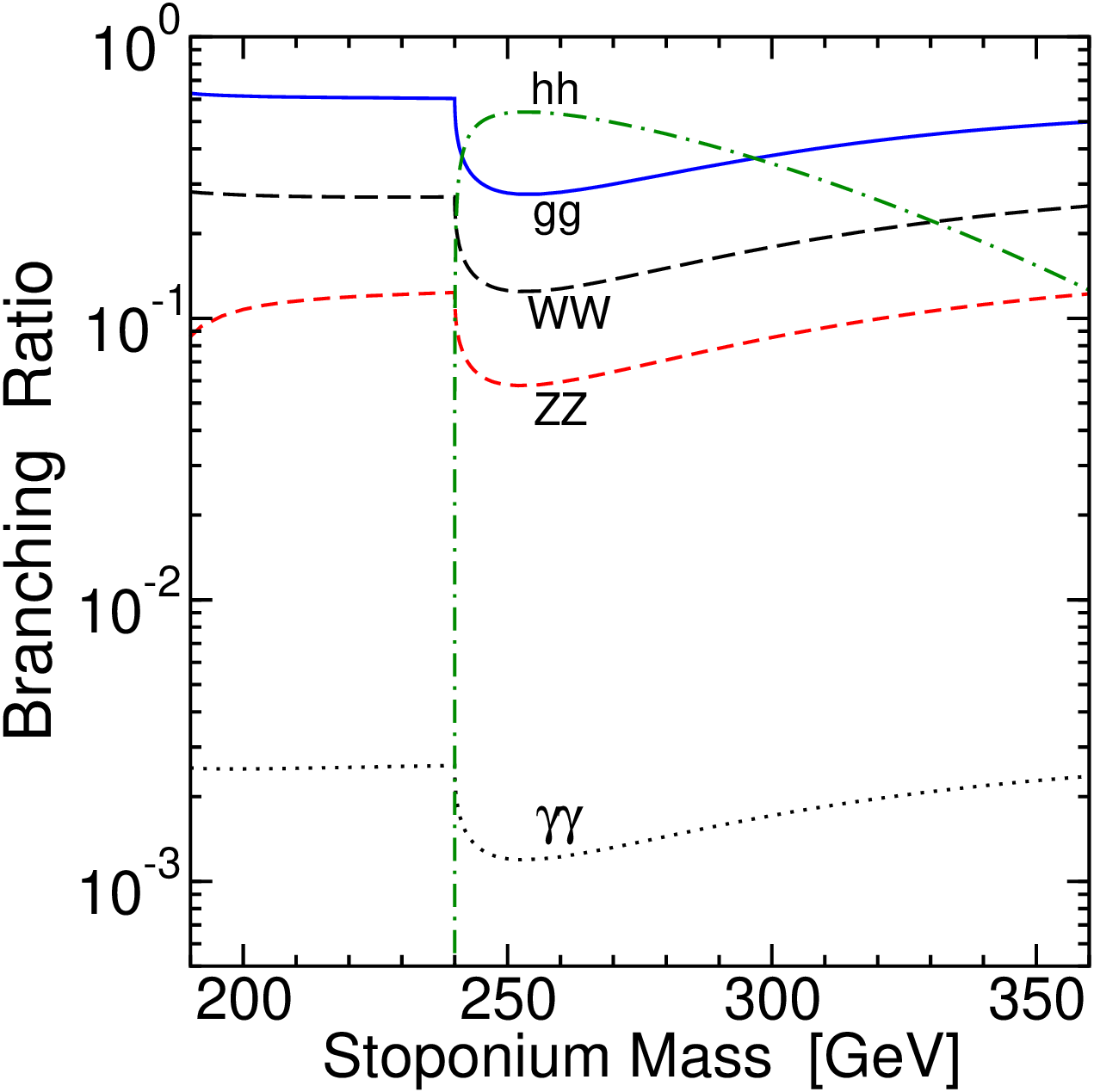}
~~\includegraphics[width=8.0cm,angle=0]{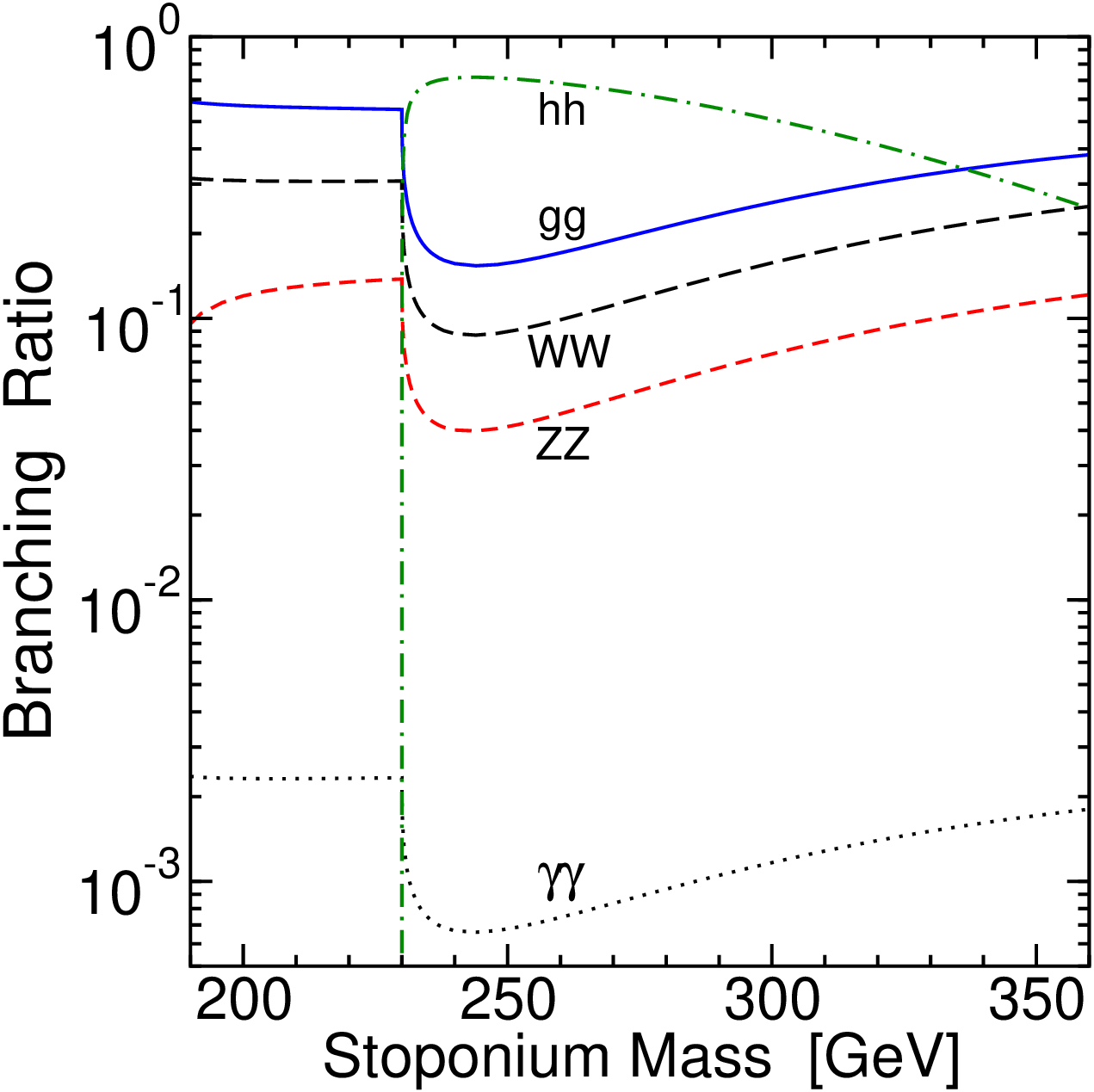}
\caption{\label{fig:baryo}
The branching ratios of scalar stoponium into $gg$, $\gamma\gamma$,
$W^+W^-$, $ZZ$, and $h^0h^0$ final states, for model lines motivated by  
electroweak-scale baryogenesis, as described in the text, with varying 
$m_{\tilde t_1}$. The left panel depicts a relatively optimistic case 
with $m_{h^0} = 115$ GeV, $|X_t|/m_{\tilde t_2} = 0.3$ and the right panel 
a pessimistic case with $m_{h^0} = 120$ GeV, $|X_t|/m_{\tilde t_2} = 
0.5$. The range that can lead to electroweak-scale baryogenesis in the
MSSM includes roughly 235 GeV $< m_{\stoponium} <$ 270 GeV.}
\end{figure}

In fig.~\ref{fig:baryo}, I show the relevant branching ratios for
stoponium decay in a relatively optimistic case with $m_{h^0} = 120$ GeV
and $|X_t|/m_{\tilde t_2} = 0.3$ (left panel), and a pessimistic case with
$m_{h^0} = 115$ GeV and $|X_t|/m_{\tilde t_2} = 0.5$ (right panel).
In both cases, $\tan\beta = 10$.
Unfortunately, the branching ratio for the decay 
$\stoponium \rightarrow h^0h^0$ is seen to be quite large 
above threshold \cite{Barger:1988sp,Drees:1993uw},
due to a small denominator (coming from the top-squark propagator)
in the last term in eq.~(\ref{eq:hhwidth}).
The BR$(\stoponium \rightarrow h^0h^0)$ 
decreases as one moves to higher stoponium masses.
I have made the optimistic but not unreasonable 
assumption that the other neutral Higgs boson $H^0$
is sufficiently heavy that the decay $\stoponium \rightarrow
b\overline b$ is not near resonance and
can be neglected. I have also optimistically assumed that
the LSP is close enough in mass to $\tilde t_1$ so that
$\stoponium \rightarrow \tilde N_1 \tilde N_1$ is unimportant
due to kinematic suppression. The existing collider lower limits
on $m_{\tilde t_1}$ from the Tevatron and LEP2 do not constrain
the top squark when $m_{\tilde t_1} - m_{\tilde N_1}$ is
small, because of the softness of the charm jets from the decay.
Obtaining a thermal dark matter density in agreement with WMAP from
stop-neutralino co-annihilations in these models requires
$m_{\tilde t_1} - m_{\tilde N_1} \gsim 20$ GeV, but a smaller mass
difference is still allowed, since it 
would just mean that the dark matter is something else, for example 
axions, or LSPs arising from a non-thermal source.

\begin{figure}[!tp]
\includegraphics[width=8.0cm,angle=0]{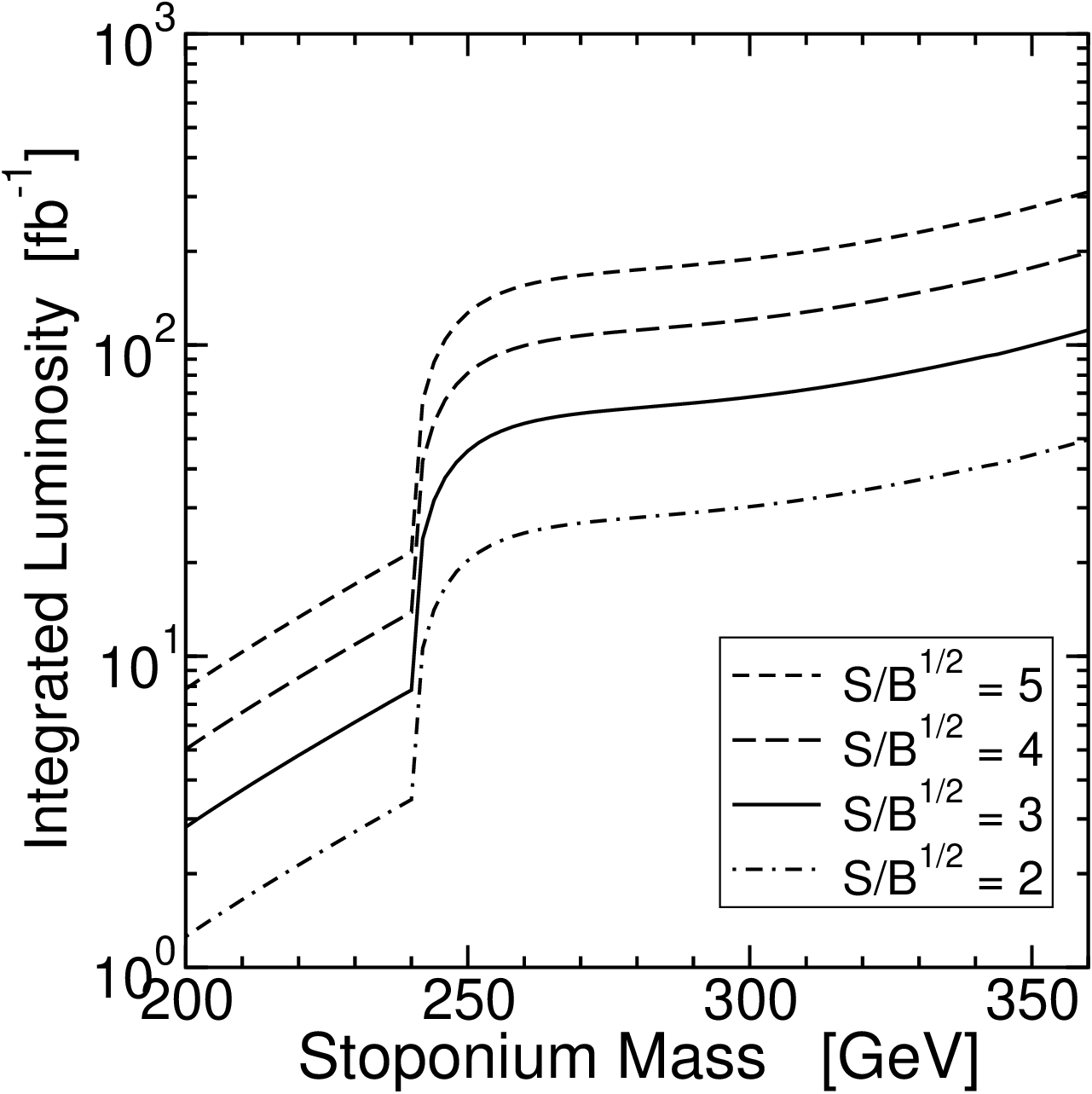}
~~\includegraphics[width=8.0cm,angle=0]{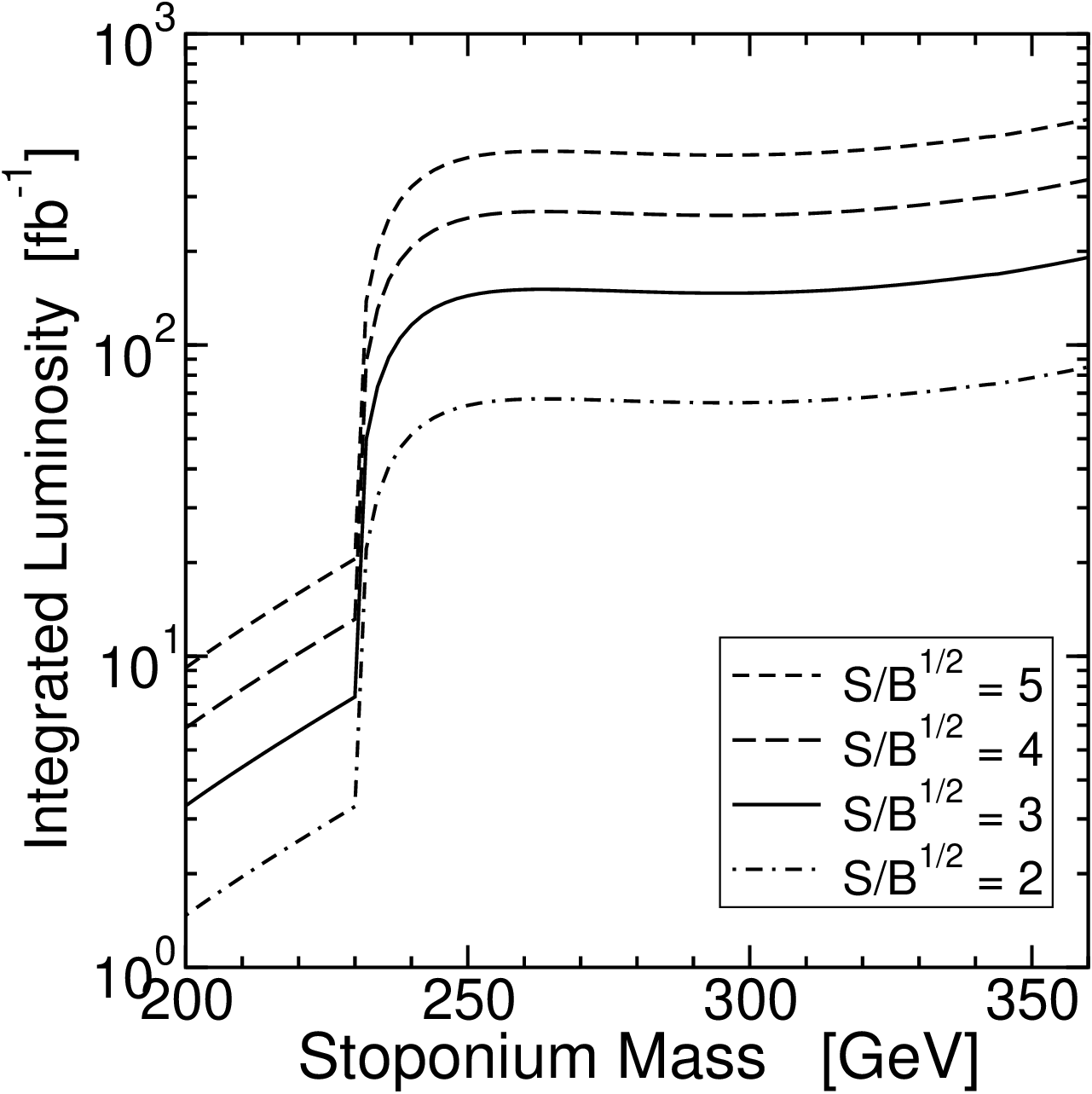}
\caption{\label{fig:baryolumin}
Total integrated luminosity yielding an expected
$S/\sqrt{B} = 2,3,4,5$ for $M_{\gamma\gamma}$ 
in a bin with $|M_{\gamma\gamma} - m_{\stoponium}|
< 0.02 m_{\stoponium}$, as a function of the stoponium mass
$m_{\stoponium}$, for $pp$ collisions at
$\sqrt{s} = 14$ TeV, for the two model lines 
depicted in fig.~\ref{fig:baryo}.
The range that can lead to electroweak-scale baryogenesis in the
MSSM includes roughly 235 GeV $< m_{\stoponium} <$ 270 GeV.} 
\end{figure}

Despite the large branching ratio to $h^0h^0$, 
there are still good prospects for stoponium
detection at the LHC in these models, because the stoponium mass
is necessarily not too large. This can be seen in 
fig.~\ref{fig:baryolumin}, which depicts the required
luminosity for detection at various expected significances, i.e. the
results of
fig.~\ref{fig:lumin} divided by  
${\rm BR}(\stoponium \rightarrow gg)]^2$.
In the most optimistic case, a clear observation of stoponium could
be possible with less than 100 fb$^{-1}$ over a large
range of stoponium masses including
the range that can accommodate electroweak-scale baryogenesis.
If the decay $\stoponium \rightarrow h^0 h^0$ is kinematically
forbidden, an observation could even be made with less than
10 fb$^{-1}$.

The detection of direct open light top squark production will probably be 
quite difficult at the LHC, especially if the $m_{\tilde t_1} - m_{\tilde 
N_1}$ mass difference is very small, because then the charm jets from 
$\tilde t_1 \rightarrow c \tilde N_1$ will be very soft. Most top squarks 
will likely come from gluino pair production followed by the decay 
$\tilde g \rightarrow t \tilde t_1$. The Majorana nature of the gluino 
implies that half of the resulting events will have like-charge top 
quarks and soft charm jets. This was found to be a viable signal in 
\cite{Kraml:2005kb}, and endpoint analyses will allow the determination 
of relations between the gluino, lighter stop, and LSP masses. A 
relatively precise measurement of $m_{\stoponium}$ would clearly be very 
helpful in pinning down the masses of all three particle.

%%%%%%%%%%%%%%%%%%%%%%%%%%%%%%%%%%%%%%%%%%%%%%%%%%%%%%%%%%%%%%%%%%%%%
\section{Outlook}
\label{sec:outlook}
\setcounter{equation}{0}
\setcounter{footnote}{1}

In this paper, I have examined prospects for observing stoponium at the 
LHC in the process $pp \rightarrow \stoponium \rightarrow \gamma\gamma$, 
as first suggested by Drees and Nojiri in 
\cite{Drees:1993yr,Drees:1993uw}. As illustrated in two distinct 
motivated scenarios, this search will likely be a long-term project, 
requiring good precision in the electromagnetic calorimetry and high 
integrated luminosity. For a positive detection with 100 fb$^{-1}$ or 
less, it is estimated to be probably necessary (but certainly not 
sufficient) that the stoponium mass is less than about 500 GeV, which 
happens to be in the middle of the range preferred by compressed 
supersymmetry models of the type discussed in 
refs.~\cite{Martin:2007gf,Martin:2007hn}. If the stoponium mass is less 
than 300 GeV, detection might even be possible with 10 fb$^{-1}$ or less. 
My estimates of the detectability for the idealized case of models for 
which BR$(\stoponium \rightarrow gg) \approx 1$ are somewhat more 
pessimistic than in \cite{Drees:1993uw}, but there is a clear opportunity 
if Nature is kind.

It would be useful to understand the higher-order corrections to 
stoponium production and decay, which might well have a large impact on 
the viability of the signal. Because these higher-order corrections are 
lacking for the signal, I have not included them for the background 
either, but they are likely to be substantial in both cases. In practice, 
the background will be obtainable from data using a sideband analysis. 
There is also an unknown, but possibly large, benefit from production of 
the excited states of stoponium adding to the signal.

An important remaining question is how accurately the observation of 
stoponium can determine the lighter top-squark mass, and from it other 
superpartner masses. The answer is quite sensitive to unknowns, including 
the true size of the backgrounds, the experimental mass resolution and 
systematic errors for diphotons, as well as an estimate of theoretical 
errors in the binding energies and wavefunctions for stoponium bound 
states. However, it seems clear that the observation of stoponium would 
present a unique opportunity to gain precise information about the 
superpartner masses, and a key ingredient in deciphering the mechanism 
behind supersymmetry breaking.

\section*{Appendix: Stoponium partial decay widths}
\label{appendixA} \renewcommand{\theequation}{A.\arabic{equation}}
\setcounter{equation}{0}
\setcounter{footnote}{1}

In this Appendix, I collect the results for the scalar stoponium 
annihilation decay widths. I agree with the results of ref. 
\cite{Drees:1993uw}, except for the $gg$ and $\gamma\gamma$ widths; 
equations (A.1) and (A.2) of that reference should each be multiplied by 
a factor of $1/2$ on the right side. (This is on top of, and distinct 
from, the factor of $1/2$ for identical particles which is correctly 
included in equation (5) of ref. \cite{Drees:1993uw}.) For these two 
widths, I agree with refs.~\cite{Moxhay:1985bg} and 
\cite{Gorbunov:2000tr}. Also, I have generalized 
ref. \cite{Drees:1993uw} slightly, by including the effects of sbottom 
mixing and possible CP violating phases. I use a convention in which 
chargino and neutralino masses are always real and positive. All 
equations below use couplings and other notations and conventions as 
given in detail in the second sections of the two papers in 
ref.~\cite{Martincons}, which will not be repeated here for the sake of 
brevity.

The general form for stoponium annihilation decay widths is
\beq
\Gamma(\stoponium \rightarrow AB) 
&=&
\frac{3 }{32 \pi^2 (1 + \delta_{AB})}
\lambda^{1/2}(1,m_A^2/m_{\stoponium}^2,m_B^2/m_{\stoponium}^2)
\frac{|R(0)|^2}{m^2_{\stoponium}}
\sum |{\cal M}|^2 ,
\eeq
for $AB =$ $gg$, $\gamma\gamma$, $ZZ$, $h^0h^0$, $Z\gamma$, $W^+W^-$, 
$t \bar t$, $b \bar b$, $\tilde N_i \tilde N_j$,
with $\delta_{AB} = 1$ for the first four cases and the last case when
$i=j$, and $\delta_{AB} = 0$ in the others.
Here 
$
\lambda(x,y,z) = x^2 + y^2 + z^2 - 2 x y - 2 x z - 2 y z ,
$
and $R(0)$ is the wavefunction at the origin, for which I 
use the $\Lambda^{(4)}_{\rm \overline{MS}} = 300$ MeV parameterization
from Table A.1 of ref.~\cite{Hagiwara:1990sq} in numerical work; see
eq.~(\ref{eq:Rzero}) of the present paper. 
It remains to give the spin-summed squared matrix element, $\sum|{\cal M}|^2$.

For the gluon-gluon and photon-photon final states, the results are 
independent of soft supersymmetry-breaking parameters, by gauge 
invariance:
\beq	
\sum |{\cal M}(\stoponium \rightarrow gg)|^2
&=& \frac{16}{9} g_3^4	
\\
\sum |{\cal M}(\stoponium \rightarrow \gamma\gamma)|^2
&=& 8 q_t^4 e^4
\eeq
where $q_t = 2/3$. 
For the $h^0h^0$ final state, 
\beq	
|{\cal M}(\stoponium \rightarrow h^0h^0)|^2
&=& \Bigl (\lambda_{h^0 h^0 \tilde t_1 \tilde t_1^*} +
\sum_{\phi^0 = h^0, H^0}  
\frac{\lambda_{\phi^0\tilde t_1 \tilde t_1^*}
\lambda_{h^0 h^0\phi^0}}{4 m^2_{\tilde t_1} - m^2_{\phi^0}}
-
\sum_{j=1,2} \frac{2 
|\lambda_{h^0 \tilde t_1 \tilde t_j^* }|^2
%\lambda_{h^0 \tilde t_j \tilde t_1^* }
}{
m^2_{\tilde t_1} + m_{\tilde t_j}^2 - m^2_{h^0}}
\Bigr )^2
.
\phantom{xxx}
\label{eq:hhwidth}
\eeq
For the $ZZ$ and $W^+W^-$ final states,
\beq	
\sum |{\cal M}(\stoponium \rightarrow ZZ)|^2
&=& 2 (a^T_{ZZ})^2 + (a^L_{ZZ})^2 ,
\\
\sum |{\cal M}(\stoponium \rightarrow W^+W^-)|^2
&=& 2 (a^T_{WW})^2 + (a^L_{WW})^2 ,
\eeq
where
\beq
a_{ZZ}^T &=& \frac{2}{g^2 + g^{\prime 2}}
\left [ 
\left (\frac{g^2}{2} - \frac{g^{\prime 2}}{6} \right )^2 
|L_{\tilde t_1}|^2
+ \frac{4g^{\prime 4}}{9}  |R_{\tilde t_1}|^2 \right ] +
\sum_{\phi^0 = h^0, H^0}  
\frac{\lambda_{\phi^0 \tilde t_1 \tilde t_1^*}
g_{ZZ\phi^0}}{4 m^2_{\tilde t_1} - m^2_{\phi^0}} ,
\\
a^L_{ZZ} &=& (1 - 2 m^2_{\tilde t_1}/m_Z^2) a^T_{ZZ}
\nonumber \\ 
&&
+ \sum_{j=1,2} 
\frac{8}{g^2 + g^{\prime 2}} \left
|\left (\frac{g^2}{2} - \frac{g^{\prime 2}}{6} \right ) 
L_{\tilde t_1} L^*_{\tilde t_j}
- \frac{2g^{\prime 2}}{3}  R_{\tilde t_1} R^*_{\tilde t_j} 
\right |^2
\frac{m^4_{\tilde t_1}/m_Z^2 - m^2_{\tilde t_1}}{
m^2_{\tilde t_1} + m^2_{\tilde t_j} - m_Z^2} ,
\\
a^T_{WW} &=& \frac{g^2}{2} |L_{\tilde t_1}|^2 +
\sum_{\phi^0 = h^0, H^0}  
\frac{\lambda_{\phi^0 \tilde t_1 \tilde t_1^*}
g_{WW\phi^0}}{4 m^2_{\tilde t_1} - m^2_{\phi^0}}
,
\\
a^L_{WW} &=& (1 - 2 m^2_{\tilde t_1}/m_W^2) a^T_{WW}
+ \sum_{j=1,2} 
2 g^2 |L_{\tilde t_1} L_{\tilde b_j}|^2
\frac{m^4_{\tilde t_1}/m_W^2 - m^2_{\tilde t_1}}{
m^2_{\tilde t_1} + m^2_{\tilde b_j} - m_W^2} .
\eeq
For the $Z \gamma$ final state,
\beq	
\sum |{\cal M}(\stoponium \rightarrow Z\gamma)|^2
&=& 
2 q_t^2 e^2 (g^2 + g^{\prime 2}) (|L_{\tilde t_1}|^2 - 4 s_W^2/3)^2 .
\eeq
The top-quark and bottom-quark final states have:
\beq	
\sum |{\cal M}(\stoponium \rightarrow t \bar t)|^2
&=& 6 (m_{\tilde t_1}^2 - m_t^2) 
(2 {\rm Re}[a_{t\bar t}] - m_t b_{t\bar t})^2
+ 24 m_{\tilde t_1}^2 ({\rm Im}[a_{t\bar t}])^2,
\\
\sum |{\cal M}(\stoponium \rightarrow b \bar b)|^2
&=& 6 (m_{\tilde t_1}^2 - m_b^2) 
(2 {\rm Re}[a_{b\bar b}] - m_b b_{b\bar b})^2
+ 24 m_{\tilde t_1}^2 ({\rm Im}[a_{b\bar b}])^2 ,
\eeq
where
\beq
a_{t\bar t} &=& 
\frac{(8 g_3^2/9)
m_{\tilde g} L^*_{\tilde t_1} R_{\tilde t_1}
}{m_{\tilde t_1}^2 + m_{\tilde g}^2 - m_t^2}
+ \frac{y_t}{\sqrt{2}}
\sum_{\phi^0 = h^0, H^0, A^0}
\frac{\lambda_{\phi^0\tilde t_1 \tilde t_1^*} k_{u\phi^0}}{
4 m_{\tilde t_1}^2 - m^2_{\phi^0}}
- \frac{1}{3} \sum_{j=1}^{4}
\frac{ m_{\tilde N_j} 
Y_{t \tilde N_j \tilde t_1^*} 
Y_{\overline t \tilde N_j \tilde t_1}}{m_{\tilde t_1}^2 
+ m_{\tilde N_j}^2 - m_t^2} 
,
\phantom{xxx}
\\
b_{t\bar t} &=& \frac{8 g_3^2/9}{m_{\tilde t_1}^2 + m_{\tilde g}^2 - m_t^2}
+ \frac{1}{3} \sum_{j=1}^{4}
\frac{ |Y_{t \tilde N_j \tilde t_1^*}|^2 + 
|Y_{\overline t \tilde N_j \tilde t_1}|^2}{m_{\tilde t_1}^2 
+ m_{\tilde N_j}^2 - m_t^2} ,
\\
a_{b\bar b} &=& 
\frac{y_b}{\sqrt{2}}
\sum_{\phi^0 = h^0, H^0, A^0}
\frac{\lambda_{\phi^0\tilde t_1 \tilde t_1^* } k_{d\phi^0}}{
4 m_{\tilde t_1}^2 - m^2_{\phi^0}}
- \frac{1}{3} \sum_{j=1}^{2}
\frac{ m_{\tilde C_j} Y_{b \tilde C_j \tilde t_1^*} 
Y_{\overline b \tilde C_j \tilde t_1}}{
m_{\tilde t_1}^2 + m_{\tilde C_j}^2 - m_b^2} ,
\\
b_{b\bar b} &=&
\frac{1}{3} \sum_{j=1}^{2}
\frac{ |Y_{b \tilde C_j \tilde t_1^*}|^2 + 
|Y_{\overline b \tilde C_j \tilde t_1}|^2}{
m_{\tilde t_1}^2 + m_{\tilde C_j}^2 - m_b^2}
.
\eeq
Finally, the final state with neutralinos has:
\beq	
\sum |{\cal M}(\stoponium \rightarrow \tilde N_j \tilde N_k)|^2
&=&
|a_{\tilde N_j\tilde N_k}|^2 
(8 m_{\tilde t_1}^2 - 2 m_{\tilde N_j}^2 - 2 m_{\tilde N_k}^2)
- 4 m_{\tilde N_j} m_{\tilde N_k} {\rm Re}[(a_{\tilde N_j \tilde N_k} )^2]
\nonumber \\ &&
+ |b_{\tilde N_j \tilde N_k}|^2 \left [
2 m_{\tilde t_1}^2 (m_{\tilde N_j}^2 + m_{\tilde N_k}^2)
- 3 m_{\tilde N_j}^2 m_{\tilde N_k}^2
- (m_{\tilde N_j}^4 + m_{\tilde N_k}^4)/2 \right ]
\nonumber \\ &&
+ {\rm Re}[(b_{\tilde N_j \tilde N_k})^2] m_{\tilde N_j} m_{\tilde N_k} (
4 m_{\tilde t_1}^2 - 2 m_{\tilde N_j}^2 - 2 m_{\tilde N_k}^2) 
\nonumber \\ &&
+ 2 {\rm Re}[a_{\tilde N_j \tilde N_k} b_{\tilde N_j \tilde N_k}]
m_{\tilde N_j} (
4 m_{\tilde t_1}^2 - m_{\tilde N_j}^2 - 3 m_{\tilde N_k}^2) 
\nonumber \\ &&
+ 2 {\rm Re}[a_{\tilde N_j \tilde N_k} b_{\tilde N_j \tilde N_k}^*]
m_{\tilde N_k} (
4 m_{\tilde t_1}^2 - 3 m_{\tilde N_j}^2 - m_{\tilde N_k}^2) ,
\label{eq:MNN}
\eeq
where
\beq
a_{\tilde N_j \tilde N_k}
&=& 
\frac{m_t (Y_{t \tilde N_j \tilde t_1^*} 
Y_{\overline t \tilde N_k \tilde t_1} +
Y_{\overline t \tilde N_j \tilde t_1}
Y_{t \tilde N_k \tilde t_1^*} 
)}{
m_{\tilde t_1}^2 + m_t^2 - (m_{\tilde N_j}^2 + m_{\tilde N_k}^2)/2}
\>\,
- \sum_{\phi^0 = h^0, H^0, A^0}
\frac{\lambda_{ \phi^0 \tilde t_1 \tilde t_1^*}
Y_{\tilde N_j \tilde N_k \phi^0 }}{
4 m_{\tilde t_1}^2 - m_{\phi^0}^2}
,
\\
b_{\tilde N_j \tilde N_k}
&=& 
\frac{Y_{t \tilde N_j \tilde t_1^*}  Y_{t \tilde N_k \tilde t_1^*}^*
+
Y_{\overline t \tilde N_j \tilde t_1}^* 
Y_{\overline t \tilde N_k \tilde t_1} 
}{
m_{\tilde t_1}^2 + m_t^2 - (m_{\tilde N_j}^2 + m_{\tilde N_k}^2)/2}
.
\eeq
If there are no CP-violating phases, then $a_{\tilde N_j \tilde N_k}$
and
$b_{\tilde N_j \tilde N_k}$ are real, and eq.~(\ref{eq:MNN}) simplifies to:
\beq
\sum |{\cal M}(\stoponium \rightarrow \tilde N_j \tilde N_k)|^2
&=&
2 [4 m^2_{\tilde t_1} - (m_{\tilde N_j} + m_{\tilde N_k})^2]
[a_{\tilde N_j \tilde N_k} 
+ (m_{\tilde N_j} + m_{\tilde N_k}) b_{\tilde N_j \tilde N_k}/2]^2.
\eeq

%%%%%%%%%%%%%%%%%%%%%%%%%%%%%%%%%%%%%%%%%%%%%%%%%%%%%%%%%%%%%%%%%%%%%%%%% 
\bigskip \noindent 
{\it Acknowledgments:} I thank Tulika Bose, Gustaaf Brooijmans, 
Manuel Drees, Germano Nardini, Mihoko Nojiri and Carlos Wagner for helpful 
communications. This work was supported in part by the 
National Science Foundation grant number PHY-0456635.

%%%%%%%%%%%%%%%%%%%%%%%%%%%%%%%%%%%%%%%%%%%%%%%%%%%%%%%%%%%%%%%%%%%%%%%%

\end{document}